\documentclass[twocolumn]{article}

\usepackage[letterpaper,top=2cm,bottom=2cm,left=3cm,right=3cm,marginparwidth=1.75cm]{geometry}
\usepackage{setspace}

\usepackage{amsmath}
\usepackage{graphicx}
\usepackage[colorlinks=true, allcolors=blue]{hyperref}
\usepackage{amsmath}
\usepackage{amssymb}
\usepackage{bm}
\usepackage[subrefformat=parens]{subcaption}

\title{
    Single-channel electroencephalography decomposition by detector-atom network and its pre-trained model
}
\author{Hiroshi Higashi\thanks{
    Graduate School of Engineering, Osaka University, Suita, Osaka, Japan.
    Phone: +81-6-6879-7745.
    Email: higashi@comm.eng.osaka-u.ac.jp}
}
\date{}

\begin{document}
\maketitle

\begin{abstract}
Signal decomposition techniques utilizing multi-channel spatial features are critical for analyzing, denoising, and classifying electroencephalography (EEG) signals.
To facilitate the decomposition of signals recorded with limited channels, this paper presents a novel single-channel decomposition approach that does not rely on multi-channel features.
Our model posits that an EEG signal comprises short, shift-invariant waves, referred to as atoms.
We design a decomposer as an artificial neural network aimed at estimating these atoms and detecting their time shifts and amplitude modulations within the input signal.
The efficacy of our method was validated across various scenarios in brain--computer interfaces and neuroscience, demonstrating enhanced performance.
Additionally, cross-dataset validation indicates the feasibility of a pre-trained model, enabling a plug-and-play signal decomposition module.
\end{abstract}


\section{Introduction} \label{sec:intro}
Signal decomposition is a pivotal technique in signal and data processing~\cite{Cichocki2002a}.
This technique finds widespread applications in areas like audio and speech~\cite{Virtanen2007}, and image~\cite{Minaee2022} processing, as well as in the analysis of electroencephalogram (EEG) signals.
There exists a wealth of techniques specifically designed to dissect EEG signals into constituent components.
These decomposition methods serve various purposes, including noise reduction~\cite{Vorobyov2002}, source separation/localization~\cite{Asadzadeh2020}, and facilitating brain--computer interfaces (BCI)~\cite{Lotte2018,Gu2021}.

EEG recordings are typically obtained using multiple electrodes in a multi-channel configuration.
Over the years, advanced signal decomposition techniques that leverage the spatial features of signals, such as beamforming~\cite{VanVeen1988} and spatial filtering, have been thoroughly developed.
For example, independent component analysis (ICA)~\cite{Hyvarinen2000} predicates the statistical independence of decomposed signals and estimates their spatial mixture.
While multi-channel measurement devices ware widely used in research and medical purposes, there is a rising trend of consumer devices with fewer channels.
However, for these devices equipped with fewer channels, wherein the electrodes may be considerably distant from one another, it is difficult to effectively leverage the spatial features for signal decomposition.

For measurement devices that capture signals with spatially-sparse electrodes, single-channel signal decomposition is often the technique of choice~\cite{Maddirala2018}.
Traditional frequency-driven methods like Fourier and (continuous) wavelet transforms~\cite{Mallat1998} are apt for this scenario.
In these decomposition approaches, decomposed signals are anchored to specific theoretical templates, such as sinusoidal signals and short waveforms, termed ``mother wavelets''.
This frequency-driven decomposition operates on the implicit assumption that distinct EEG components segregate into separate frequency bands.
However, when it comes to EEG decomposition, components might not be neatly categorized into specific frequency bands; they can span and overlap across multiple frequency bands~\cite{Munck2009,Babadi2014}.

An alternative to single-channel signal decomposition is the adoption of data-driven or heuristic techniques---known for their flexibility in decomposing signals~\cite{Bajaj2012}.
A notable example is empirical mode decomposition (EMD)~\cite{Huang1998}, which has found applications in EEG analsysis~\cite{Sweeney2018}.
Many prevalent data-driven methods aim for adaptive application, meaning they rely solely on the signal-to-be-decomposed for the design of the decomposition.
However, a potential drawback is that when the input signal is relatively short, its decomposition can be significantly influenced by noise.
Furthermore, the decomposition lacks consistency across multiple observed signals, making it challenging to determine whether a decomposed signal is desired or not.

As we introduced earlier, EEG decomposition has traditionally employed prior knowledge encompassing spatial features, frequency assumptions, and statistical hypotheses to facilitate robust decomposition.
Here, we propose a new data-driven approach that capitalizes on training data, not relying on the prior knowledge.
In most practical situations where EEG decomposition is needed, such data that we can use for training a decomposer is available.
For instance, in the realm of neuroscience, EEG data from numerous trials is often aggregated and subjected to a collective analysis.
In a similar vein, when it comes to BCI applications~\cite{Gu2021}, training data may exist that aid in the supervised training of classifiers.
Our study is underpinned by the aspiration to harness these EEG data, convinced that such an approach holds promise for a more robust and consistent decomposition framework.

Using the training data, our proposed decomposition identifies short-length waves that collectively reconstruct the original EEG signals.
This approach is based on the model proposed by Brockmeier and Principe~\cite{Brockmeier2016}, which suggests the existence of recurrent wave patterns within EEG signals.
Theoretically, key EEG signatures, such as oscillatory patterns---including alpha, mu, beta, and gamma rhythms~\cite{Jas2017}---as well as event-related potentials (ERPs)~\cite{Barthelemy2013}, can be modeled by these recurrent waves.
The proposed decomposer estimates recurrent waves, akin to shift-invariant \textit{atoms} in dictionary learning frameworks~\cite{Kreutz2003,Grosse2007,Simon2019,Garcia2018}, using either supervised or unsupervised data-driven methods.
Analogous to continuous wavelet transform techniques~\cite{Unser1996}, our method reconstructs signals by overlaying time-shifted and amplitude-modulated atoms.
The time shifts and amplitude modulations of the atoms are identified by neural network modules, which we refer to as \textit{detectors}.
This reconstruction process is formulated as a convolution of the detector outputs with the atoms, implemented via convolutional layers~\cite{Tao2021,Stankovic2023,Davies2024}.
By employing an artificial neural network framework, we enable optimization of the detectors and atoms in the decomposer using various loss functions, including those derived from supervised learning principles.
This approach supports the flexible design of signal models tailored to distinct EEG signatures through the development of specialized neural network architectures.

In this paper, we introduce several loss functions and network architectures designed to tackle various challenges commonly encountered in neuroscience and BCI research.
A series of experiments have been conducted to verify the efficacy of these proposed losses and architectures.
These experiments demonstrated the practical utility and adaptability of our proposed decomposition technique in real-world applications.
Furthermore, we explored the feasibility of pre-training to develop a decomposer that requires no calibration.
This approach involves pre-training the network (detectors and atoms) with large datasets.
Our BCI experiment, which included a cross-dataset experimental scenario for validating dataset shifts~\cite{Quinonero2009,Dockes2021,Mellot2023}, suggests the feasibility of applying a pre-trained decomposer to any EEG signal.
We have made this pre-trained decomposer publicly available; it require only a single-channel input, thereby simplifying measurement and analysis procedures.
This advancement enables researchers, engineers, medical professionals, and consumers to decompose EEG signals in a plug-and-play manner.

In this paper, our contributions encompass:
\begin{enumerate}
    \item Introducing a robust single-channel EEG decomposition using a convolution model with a limited number of atoms.
    \item Providing empirical evidence through our experiments that demonstrates the feasibility of decomposing and reconstructing EEG signals using our model.
    \item Releasing a pre-trained decomposer to the public, enabling users to decompose EEG signals in a plug-and-play manner.
\end{enumerate}

\section{Detector-atom network}
In this paper, we introduce a novel signal decomposition method that leverages training data and artificial neural networks.
This section outlines the foundation of our decomposer network.
We begin by detailing the key components and structure of the network.
Subsequently, we discuss both unsupervised and supervised loss functions, which are employed to optimize the network.
Additionally, we introduce a technique called ``atom reassignment'' designed to prevent the optimization process from becoming stagnant.

\subsection{Signal model and network architecture} \label{sec:mm:model}
Our decomposition proposition hinges on the idea that an EEG signal can be reconstructed by a linear combination of short signals, referred to as atoms.
This concept echoes the tenets of the Fourier transform, where atoms are complex sinusoidal signals, and the wavelet transform, where atoms are mother wavelets.
What sets our method apart is the unsupervised/supervised design of these atoms.
In our approach, we employ a network comprising of detector and atom parts.
The detector is responsible for generating coefficients for linear combinations, i.e., it determines how to convolve an atom.
On the other hand, the atom network holds a single atom.
The whole network produces the decomposed signal which is generated by a convolution of the atom.

To elucidate our decomposition framework, let us delve into a simple illustration.
Consider a simple reconstruction model that decomposes a single-channel signal into multiple signals and then reconstructs the original signal by summing the decomposed signals.
We hypothesize that EEG signals can be reconstructed using $N$ atoms, implying that an EEG signal is decomposed into $N$ signals.
This model is formulated as~\cite{Lewicki1999,Grosse2007}
\begin{equation}
    x[i] = \sum_{n=1}^{N} \sum_{j=0}^{M_{n}-1} a_{n}[j] z_{n}[i - j],
    \label{eq:x_model}
\end{equation}
where
$x[i]$ is the observed signal at time index $i$,
$\{ a_{n}[j] \}_{j=0}^{M_{n}-1}$ denotes the $n$th atom,
$M_{n}$ is the length of the $n$th atom,
and
$z_{n}[i]$ is the magnitude of the $n$th atom at time index $i$.
Given that $z_{n}[i]$ represents the magnitudes, $z_{n}[i]$ must uphold a non-negative constraint($z_{n}[i] \geq 0$).
The objective of our proposed networks is to estimate the atoms $a_{n}$ and their respective magnitudes $z_{n}$.

\begin{figure}[tp]
    \centering
    \includegraphics[width=.95\linewidth]{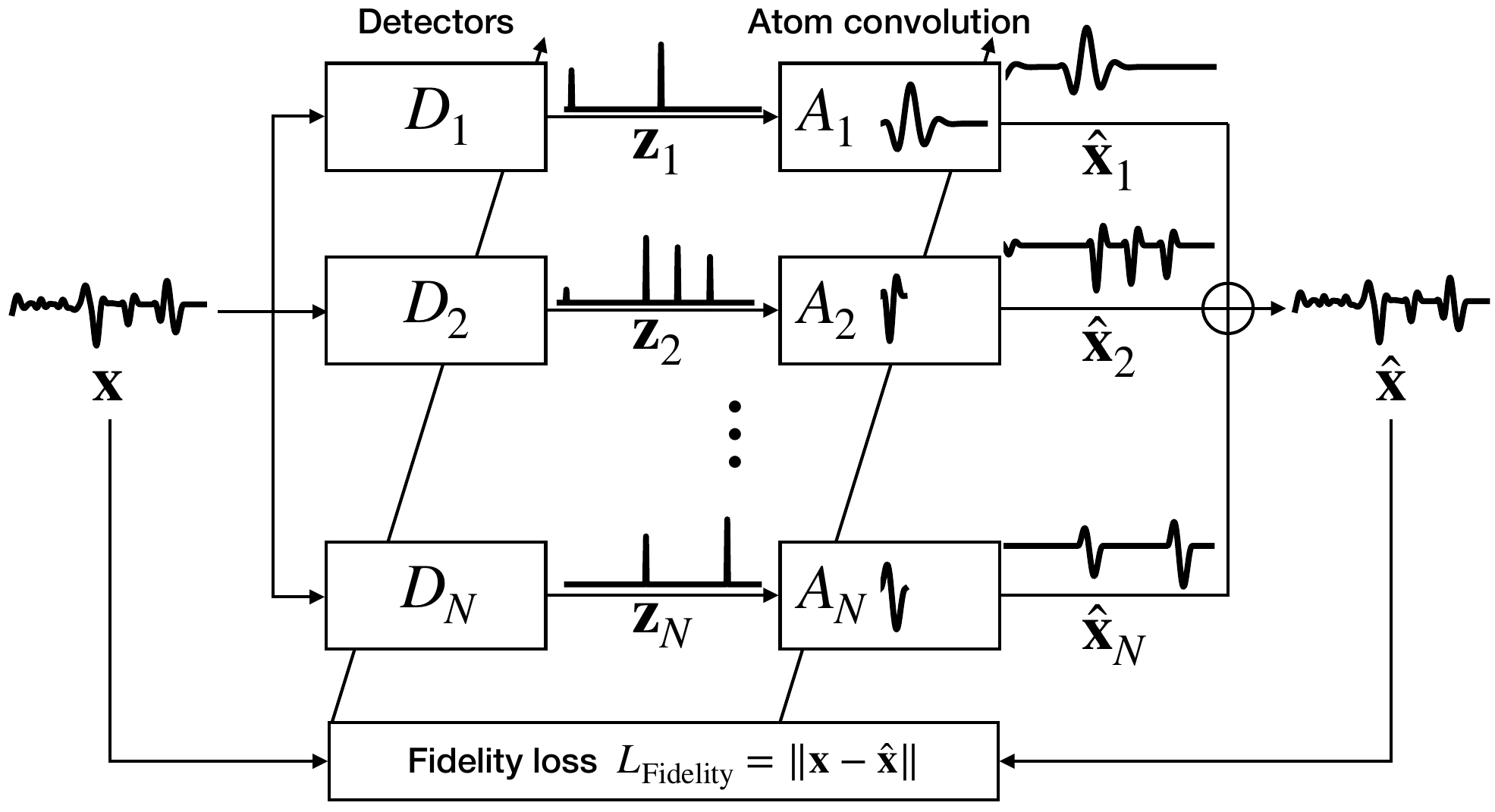}
    \caption{
        Schematic of the fundamental detector-atom network architecture.
    }
    \label{fig:fundam}
\end{figure}

The integrated detector-atom network aims to determine these components using training data.
The fundamental architecture of this network is visually represented in Figure~\ref{fig:fundam}.
The detector's primary responsibility is deducing $z_{n}[i]$.
Any network that can generate a non-negative signal, congruent in length to the input signal $x[i]$, can potentially function as a detector.
For simple notation, we opt for a solitary convolutional layer, and show an implementation example.
The output of the $n$th detector is formulated as
\begin{equation}
    z_{n}[i]
    = {\rm ReLU} \left( \sum_{j=0}^{P_{n}-1} d_{n}[j] x[i - j] \right),
    \label{eq:z_n}
\end{equation}
where
$x[i]$ is the observed signal and the input for the detector,
the set $\{ d[j] \}_{j=0}^{P_{n}-1}$ encompasses coefficients for the convolutional layer,
and
$P_{n}$ is the filter size of the convolutional layer.
To convey this in vector form, with a signal length of $T$, it can be represented as
\begin{equation}
    \bm{z}_{n} = D_{n}(\bm{x}),
\end{equation}
where
$\bm{x}, \bm{z}_{n} \in \mathbb{R}^{T}$ are the vector forms of the observed signal and the $n$th detector output, defined respectively as $\bm{x} = [x[0], x[1], \ldots, x[T - 1]]^{\top}$ and $\bm{z}_{n} = [z_{n}[0], z_{n}[1], \ldots, z_{n}[T - 1]]^{\top}$, respectively, and
$D_{n} \colon \mathbb{R}^{T} \to \mathbb{R}_{+}^{T}$ is an operator, as described in (\ref{eq:z_n}).
Both $\bm{x}$ and $\bm{z}_{n}$ maintain congruent signal lengths.
Owing to the non-negative activation function, the detector's output results in non-negative values.
The detector's output is conceived to, within the input signal $x[i]$, identify the time shift and amplitude modulation of the atoms, which is delineated by the subsequent atom network $A_{_n}$.

The subsequent convolution with the atoms produces a decomposed signal:
\begin{equation}
    \hat{x}_{n}[i] = \sum_{j=1}^{M_{n}} a_{n}[j] z_{n}[i - j].
    \label{eq:x_n}
\end{equation}
In vector form, the signal reconstruction by the $N$ atom network outputs can be described as
\begin{equation}
    \hat{\bm{x}}_{n} = A_{n}(\bm{z}_{n}),
\end{equation}
where 
$\hat{\bm{x}}_{n} \in \mathbb{R}^{T}$ is the vector of the reconstructed signal, formulated as $\hat{\bm{x}} = [\hat{x}_{n}[0], \hat{x}_{n}[1], \ldots, \hat{x}_{n}[T - 1]]^{\top}$,
and
$A_{n} \colon \mathbb{R}^{T} \to \mathbb{R}^{T}$ is an operator, defined in accordance with (\ref{eq:x_n}).

If we interpret the atoms as elements of a dictionary and the detector outputs as (sparse) representations using those elements, we found that this decomposition procedure is fundamentally the same as convolutional dictionary learning (CDL)~\cite{Garcia2018}.
Typically, CDL optimizes the dictionary and representations to minimize reconstruction loss.
In contrast, our proposed method, which implements these dictionary and representations as neural network modules, can use flexible optimization strategies tailored to EEG characteristics and the specific goals of the decomposition, as we will introduce in the next section.

\subsection{Loss for network optimization} \label{sec:mm:loss}
The parameters subject to optimization encompass the coefficient set for a detector and an atom in each detector-atom network:
$ \{
\{ \bm{d}_{1}, \bm{a}_{1} \},
\{ \bm{d}_{2}, \bm{a}_{2} \},
\ldots,
\{ \bm{d}_{N}, \bm{a}_{N} \}
\} $,
where $\bm{d}_{n} = [d_{n}[0], d_{n}[1], \ldots, d_{n}[P_{n} - 1]]^{\top}$
and $\bm{a}_{n} = [a_{n}[0], a_{n}[1], \ldots, a_{n}[M_{n} - 1]]^{\top}$
for $n = 1, \ldots, N$.
The objective of this optimization is to minimize the discrepancy between the observed signal $x[i]$ and the reconstructed signal $\sum_{n=1}^{N} \hat{x}_{n}[i]$.
This discrepancy is quantified as
\begin{equation}
    L_{\rm Fidelity} = \left\| \bm{x} - \sum_{n=1}^{N} \hat{\bm{x}}_{n} \right\|.
    \label{eq:fidelity}
\end{equation}
The loss function $L_{\rm Fidelity}$ is minimized throughout the training phase.
Consequently, the detector-atom network is trained to decompose the input observed signal and accurately reconstruct the input from the decomposed ones, each of which is represented by an atom.

Given labeled training samples, meaning pairs of signals and their corresponding class labels as $(\bm{x}, y)$, it is possible to define supervised loss functions tailored to the class.
Suppose we have signals assigned to $N_{c}$ distinct classes, represented as $\omega_{1}, \omega_{2}, \ldots, \omega_{N_{c}}$.
Our approach is to design $N_{c}$ individual decomposers.
Each decomposer is specialized; its primary function is to decompose and reconstruct signals that belong to a specific class.
This concept is formulated by a supervised (SV) loss function for the $c$th decomposer;
\begin{equation}
    L^{(c)}_{\rm SV} =
    \left\{
    \begin{array}{ll}
        \| \bm{x} - \tilde{\bm{x}}_{c} \| & \quad (y = \omega_{c}), \\
        \| \tilde{\bm{x}}_{c} \| & \quad (\mathrm{otherwise}),
    \end{array}
    \right.
    \label{eq:loss:supervised}
\end{equation}
for $c = 1, 2, \ldots, N_{c}$,
where $\tilde{\bm{x}}_{c}$ represents the sum of signals decomposed by the $c$th decomposer.
If the signal class $y$ matches the class for which the decomposer is designed, the loss is the difference between the observed signal and the decomposed one.
If the signal class $y$ does not match, the loss is simply the magnitude of the reconstructed signal.
This ensures that the decomposer specific to other classes does not attempt to reconstruct the signal that does not belong to its designated class.

Promoting sparsity in the dictionary representation helps prevent overfitting and enhances robustness~\cite{Chen1998}.
For our proposed decomposer, we introduce a sparsity loss for the outputs of the detectors, defined by
\begin{equation}
    L_{\rm Sparsity} = \sum_{n=1}^{N} \| \bm{z}_{n} \|_{1}.
    \label{eq:l_sparse}
\end{equation}
This loss is combined with a fidelity loss, $L_{\rm Fidelity}$ or $L_{\rm SV}$, to form a total loss function such as $L = L_{\rm Fidelity} + \alpha_{\rm Sparse} L_{\rm Sparse}$,
where $\alpha_{\rm Sparsity}$ is a regularization coefficient.

When promoting sparsity in detector outputs, pairs of detector outputs and atoms often tend to converge toward zero at the beginning of training, before the reconstruction performance is still low.
Once this zero convergence occurs, these pairs do not produce any non-zero decomposed signals in later epochs and thus do not contribute to improving the reconstruction loss---they effectively become deactivated.
To explore the potential contribution of these deactivated pairs to reconstruction performance in the later optimization phase, when reconstruction performance is relatively high, we propose a technique called {\it atom reassignment}.
Atom reassignment revives deactivated pairs by transferring the detector and a part of the atom from an active pair.
See Appendix Section~\ref{sec:atom_reassign} for the details.

\section{Applications}
The network architecture for our decomposer can be easily modified, improved, and applied, thanks to recent advances in artificial neural network.
Here, we propose several networks specifically designed to address common problems encountered in neuroscience and BCI research.

\subsection{Noise reduction using noisy-inducing event periods} \label{sec:mm:noise}
This section expands the detector-atom network-based decomposer to a noise reduction techniques tailored for scenarios where we know the timings of noise-inducing events in training samples, but not their specific nature.
For instance, consider a setup where both EEG and electrooculogram (EOG) signals are available.
The EOG signal can pinpoint the timings of ocular activities like blinks and saccades, which often induce noise into the EEG signal.
Traditional methods for noise reduction such as principal component analysis (PCA) or ICA, work under the presumption that EEG noise profiles are congruent to the EOG signal~\cite{Nolan2010}.
However, our proposed method deviates from this convention, offering a more versatile approach.
Essentially, our method is applicable with any noise reference, as long as the event timings inducing EEG noise can be ascertained.
This flexibility means it can accommodate varied noise references, including eye trackers or motion capture systems.

The network architecture for noise reduction is based on a signal observation model where the EEG signal, $x[i], i = 0, 1, \ldots, T - 1$, is a composite of brain-related signals, $s_{l}[i]$, and noise-event-related signals, $n_{l}[i]$.
This relationship is formulated as:
\begin{equation}
    x[i] = s[i] + n[i] = \sum_{l=1}^{N_{s}} s_{l}[i] + \sum_{l=1}^{N_{n}} n_{l}[i],
\end{equation}
where
$N_{s}$ is the number of brain-related signals
and $N_{n}$ is the number of noise-related signals.
The objective of this noise reduction process is the accurate estimation of the signals, $\{s_{l}[i]\}_{l=1}^{N_{s}}$ and $\{n_{l}[i]\}_{l=1}^{N_{n}}$.

The network encompasses two integral detector-atom networks:
one tailored for the extraction of brain-related signals,
and another focused on isolating signals associated with noise events.
In this setup,
$\{D_{{\rm S},l}\}_{l=1}^{N_{s}}$ and $\{A_{{\rm S},l}\}_{l=1}^{N_{s}}$ constitute the detectors and atoms for the first network, the signal-estimator, which is designed to extract brain-related signals.
Conversely, $\{D_{{\rm N},l}\}_{l=1}^{N_{n}}$ and $\{A_{{\rm N},l}\}_{l=1}^{N_{n}}$ constitute the detectors and atoms of the second network, the noise-estimator, which is designed to extract the noise-event-related signals.

The noise-event-related signal $\hat{\bm{n}} \in \mathbb{R}^{T}$ is estimated by summing the outputs of processing the input signal, $\bm{x} \in \mathbb{R}^{T}$, through each pair of detector-atom network in the noise-estimator, expressed as:
\begin{equation}
    \hat{\bm{n}} = \sum_{l=1}^{N_{m}} A_{{\rm N},l}(D_{{\rm N},l}(\bm{x})) = N_{\rm N}(\bm{x}).
\end{equation}
where $N_{\rm N}$ represents the integrated operation of the detector-atom pairs in the noise-estimator.
Following the computation of $\hat{\bm{n}}$, this estimated noise component is deduced from the original input signal $\bm{x}$, resulting in a version devoid of noise components.
Subsequently, this noise-reduced signal undergoes processing through each pair of detector-atom within the first network, culminating in the estimation of the brain-related signal $\hat{\bm{s}} \in \mathbb{R}^{T}$, articulated as:
\begin{equation}
    \hat{\bm{s}} = \sum_{l=1}^{N_{s}} A_{{\rm S},l}(D_{{\rm S},l}(\bm{x} - \hat{\bm{n}})) = N_{\rm S}(\bm{x} - \hat{\bm{n}}),
\end{equation}
where $N_{\rm S}$ represents the integrated operation of the detector-atom pairs in the signal-estimator.

Optimizing the parameters for the detector-atom pairs involves minimizing specific loss functions.
For the signal-estimator, $N_{\rm S}$, the loss function, $L_{\rm S}$, measures the deviation between the input signal and the signal combined the outputs of the signal and noise-estimators.
This is computed as
\begin{equation}
    L_{\rm S}
    = \sum_{i} | x[i] - (\hat{s}[i] + \hat{n}[i]) |_{2}^{2}
    = \| \bm{x} - (\hat{\bm{s}} + \hat{\bm{n}}) \|,
    \label{eq:signal}
\end{equation}
where
$\hat{s}[i]$ and $\hat{n}[i]$ are the $i$th time instance of $\hat{\bm{s}}$ and $\hat{\bm{n}}$.
Conversely, the loss function for the noise-estimator, $N_{\rm N}$, denoted as $L_{\rm N}$, accounts for two factors: the deviation between the input signal and the output of the noise-estimator during noisy events, and the energy of the estimated signal when no noisy event are present.
With $\mathcal{P}$ which represents the set of time indices at which noise events occur, the loss $L_{\rm N}$ is articulated as:
\begin{equation}
    L_{\rm N} = \sum_{i \in \mathcal{P}} | x[i] - \hat{n}[i] |_{2}^{2} + \sum_{i \notin \mathcal{P}} | \hat{n}[i] |^{2}.
    \label{eq:noise}
\end{equation}
This formulation ensures a precise estimation of noise-event-related signals during periods of noise events while maintaining the integrity of brain-related signal estimations outside of these periods.

Nevertheless, this approach does not consider the output of the signal-estimator, $\hat{\bm{s}}$, for designing the noise-estimator.
Thus, following the convergence of $L_{\rm S}$ or after a predetermined number of iterations, a refined loss function $L_{\rm N}^{\prime}$, is proposed.
This modification not only accounts for the noise-event-related signals but also integrates the estimated brain-related signals, formulated as:
\begin{equation}
    L_{\rm N}^{\prime} = \sum_{i \in \mathcal{P}} | x[i] - (\hat{n}[i] + \hat{s}[i]) |^{2} + \sum_{i \notin \mathcal{P}} | \hat{n}[i] |^{2}.
    \label{eq:noise2}
\end{equation}

\subsection{SSVEP detection} \label{sec:mm:ssvep}
The subsequent application focuses on SSVEP-based BCIs.
These BCIs are engineered to estimate, from an EEG signal, the target stimulus a user is fixating on.
Typically, this estimator is designed utilizing supervised methods, given the availability of training samples.
Each sample for training comprises an EEG signal paired with a class label that corresponds to the target SSVEP stimulus.

The network devised for SSVEP detection is anchored in a generative model that characterizes SSVEP responses.
Within the paradigm of an SSVEP-based BCI, visual stimuli are periodically presented at a specific frequency.
The BCI detects the corresponding EEG activity that matches the stimulus frequency.
These detected responses are indicative of the user's visual focus and are represented by a series of visual-evoked potentials corresponding to each stimulus flash.

The mathematical model of the response against a stimulus flashing at a frequency $f$ [Hz] ($\tau = 1/f$ [s]) and a phase $\phi$ [s] is defined as
\begin{equation}
    x(t) = \sum_{l=0, 1, \ldots} g_{l} v \left(t - l \tau + \phi \right) + \eta(t),
    \label{eq:ssvep}
\end{equation}
where $v(t)$ denotes the response evoked by a single flash of the stimulus,
$g_{l}$ represents the amplitude of the response to the $l$th flash,
and
$\eta(t)$ includes various forms of noise such as background EEG activity, artifacts, and responses to non-target stimuli.
For a practical scenario, we implement in a discrete-form with $x[i]$.

The network for SSVEP detection is structured with $N$ detectors, denoted as $\{D_{l}\}_{l=1}^{N}$, and a single atom network $A$.
Here, $N$ corresponds the number of distinct SSVEP stimuli in the BCI can detect.
The detectors' role is to ascertain the temporal shifts and amplitudes of the visual-evoked potentials (VEPs).
Specifically, the detectors determine the time-varying VEP amplitude $g_{l}$ in (\ref{eq:ssvep}).
These detectors are linked to a single atom network $A$, which is tantamount to the VEPs elicited by a single stimulus flash $v(t)$.
The decomposer reproduces the observed signal by
\begin{equation}
    \hat{\bm{x}} = \sum_{l=1}^{N} A(D_{l}(\bm{x})).
\end{equation}
The architecture is designed such that each detector $D_{l}$ pinpoints the temporal offset and amplitude of VEPs induced by the $l$th class of SSVEP stimulus, while the atom network $A$'s atom replicates the fundamental VEP response to the flashes across all SSVEP stimuli.

The supervised loss defined in (\ref{eq:loss:supervised}) is applied for optimizing the detector and atom networks.
The loss function is designed to measure the discrepancy between the observed EEG signal and the reconstructed signal estimated by the associated with the target SSVEP class.
Given a training sample represented as $\{\bm{x}, y\}$,
where $y \in \{1, \ldots, N\}$ is a class label corresponding to one of the $N$ possible SSVEP stimuli.
The loss function is formulated as the sum of two distinct loss components:
\begin{equation}
    L = L_{1} + L_{2},
    \label{eq:ssvep_loss}
\end{equation}
where $L_{1}$ captures the reconstruction error, quantifying the difference between the input signal and the aggregate of the signals reconstructed by all detectors:
\begin{equation}
    L_{1} =  \left\| \bm{x} - \sum_{l=1}^{N} A(D_{l}(\bm{x})) \right\|,
\end{equation}
and $L_{2}$ is a supervised loss term that penalizes the contributions of non-target class decompositions, ensuring that the reconstruction primarily utilizes the components corresponding to the target class:
\begin{equation}
    L_{2} = \sum_{l=1, l \neq y}^{N} \left\| A(D_{l}(\bm{x})) \right\|.
\end{equation}
The goal of this loss function is to ensure that the reconstructed signal is derived predominantly from the detector-atom network corresponding to the target class, thereby enhancing the specificity and accuracy of the SSVEP signal reconstruction.

Importantly, the optimization of the networks does not necessitate explicit knowledge of the stimulus frequencies and phases, which are denoted as $f$ (or $\tau$) and $\phi$ in (\ref{eq:ssvep}).
Instead, these stimulus properties are intrinsically discerned by the detectors via the class label associated with each SSVEP stimulus.

\subsection{Time-locked component extraction} \label{sec:mm:time-locked}
This application focuses on the decomposition of event-locked signals, which are time-aligned with specific event occurrences (epoch).
Event (Time)-locked signals are characterized by ERPs, forming by positive or negative fluctuations at certain latencies post-event.
These ERPs are considered to be indicative of underlying cognitive processes, exhibiting amplitude variations that correlate with these processes.
The goal here is to effectively decompose these components by introducing a constraint in the network architecture that prevents the time-shifting of atoms.

In contrast to the SSVEP decomposer delineated in Section~\ref{sec:mm:ssvep}, which utilizes a shared atom for all detectors, the ERP decomposer is designed differently to capture diverse waveforms associated with ERPs, as follows.
First, atoms are distinct across each detector, enabling the representation of the diverse ERP waveforms.
Second, the temporal shift, dictated by the detector, remains constant to extract the event-locked components.
Last, the amplitude of a detector's output quantifies the influence of the event-locked component that is triggered post-event.
The contrast bears resemblance to that between the principles of dictionary learning~\cite{Kreutz2003}, which underpin the ERP model, and convolutional dictionary learning~\cite{Garcia2018}, which is analogous to the foundational detector-atom network structure.

The network architecture for the ERP decomposer is based on a generative model~\cite{Kotani2024}.
In this model, we assume that $N$ ERPs are overlaid in an observed signal.
The waveform of the $n$th evoked potential is represented by $a_{n}[i]$.
The epoched, observed signal is then modeled as
\begin{equation}
    x[i] = \sum_{n=1}^{N} d_{n} a_{n}[i] + \eta[i], \quad i = 0, \ldots, T - 1,
\end{equation}
where $d_{n} \geq 0$ represents the amplitude of the $n$th evoked potential and $\eta[i]$ is noise.
This model assumes that the waveform $a_{n}[i]$ does not shift over time within the epoch, meaning that the potential fluctuation is time-locked.
While the atoms, $a_{1}[i], \ldots, a_{n}[i]$, do not change across trial (epoch), the corresponding amplitude, $d_{1}, \ldots, d_{N}$, vary for each epoch.
In our decomposer network, the detector functions as an estimator of these trial-by-trial amplitude variations.

The architecture of the network is composed of $N$ detectors, each corresponding to a distinct ERP.
Given that the exact number of ERPs is typically not known a priori, it is often determined through empirical estimation.
The network reconstructs the observed input signal by
\begin{equation}
    \hat{\bm{x}} = \sum_{n=1}^{N} \hat{\bm{x}}_{n} = \sum_{n=1}^{N} D_{n}(\bm{x}) \bm{a}_{n},
    \label{eq:model_erp2}
\end{equation}
where
$\hat{\bm{x}}_{n}$ is the $n$th decomposed signal given by $\hat{\bm{x}}_{n} = D_{n}(\bm{x}) \bm{a}_{n}$,
$D_{n} \colon \mathbb{R}^{T} \to \mathbb{R}_{+}$ is the $n$th detector to estimates the amplitude of the $n$th components, $d_{n}$, in the observed signal, and
the $n$th atom $\bm{a}_{n}$ serves to model the waveform associated with an event-locked ERP, $a_{n}[i]$.
This model assume that the observed signals are linear combinations of $N$ vectors, $\{\bm{a}_{1}, \ldots, \bm{a}_{N} \}$.
The unknown parameters of this model are the $N$ pairs of the detector and atom, $\{D_{n}, \bm{a}_{n}\}_{n=1}^{N}$.
To optimize these parameters, we can use the fidelity loss $L_{\rm Fidelity} $ as defined in~\eqref{eq:fidelity}.

\section{Experimental validation} \label{sec:exp}
Signal decomposition techniques hold significant potential for advancing research in neuroscience and the development of BCIs.
This section showcases toy examples of EEG signal decomposition utilizing the method proposed in this study.
These examples serve to demonstrate the practical application and effectiveness of our decomposition approach in scenarios for real-world EEG data analysis.
To validate that the proposed method can decompose an EEG signal into meaningful components, we employed the simples possible methods for other procedures, including preprocessing, feature extraction, and classification.

\subsection{Reconstruction} \label{sec:exp:reconst}
Our initial experiment focused on assessing the accuracy of signal reconstruction post-decomposition.
To evaluate the accuracy, we divided a dataset into two parts: one for training the decomposer and the other for testing.
The decomposer was trained using the training set, after which it decomposed signals from the testing set.
We then evaluated the reconstruction accuracy by comparing the original signals with those reconstructed from their decomposed counterparts.
Furthermore, to explore the potential for transfer learning, we employed datasets from different subjects, tasks, and channels as testing samples.

\subsubsection{Dataset} \label{sec:exp:reconst:dataset}
We used an open dataset~\cite{Dzianok2022} containing EEG data from 42 subjects performing four different tasks.
Due to the absence of trigger data in the recordings, the whole trials from the two subjects were excluded.
These subjects engaged in four distinct tasks: simple reaction time (SRT), Oddball, multi-source interference task (MSIT), and Rest.
While the EEG signals were originally captured using a 123-channel setup at a sampling frequency of 1000~Hz, our experiment focused primarily on the data from channel Cz for training the network.
Additionally, we utilized data from a select few other channels for testing purposes.
On average, a single session, where a subject performed a certain task, contained about 431 trials (samples).

\subsubsection{Architecture} \label{sec:exp:reconst:network}
The decomposer was based on the fundamental network architecture outlined in Section~\ref{sec:mm:model}.
We experimented with decomposers configured to generate various numbers of decomposed signals, specifically 2, 4, 8, 16, and 32 outputs.

A detector was structured with three layers, each comprising a convolutional layer followed by a ReLU activation function.
The convolutional layers were designed with a kernel size of 100, corresponding to a filter size of approximately 0.1~s, and were configured to have a single channel.
This kernel size was also adopted for the convolutional layer in the atom network.

For optimizing the decomposer, we employed an unsupervised loss function as defined in (\ref{eq:fidelity}).
The optimization process utilized an Adam optimizer with the following parameters: a learning rate of 0.001, $\beta_{1}$ of 0.5, $\beta_{2}$ of 0.999, and a weight decay of $10^{-5}$.
The optimization was conducted 1,000 iterations (epochs), with batches of 100 samples each.

\subsubsection{Result} \label{sec:exp:reconst:result}
\begin{figure}[tp]
    \centering
    \subfloat[For each task]{
        \includegraphics[width=.46\linewidth]{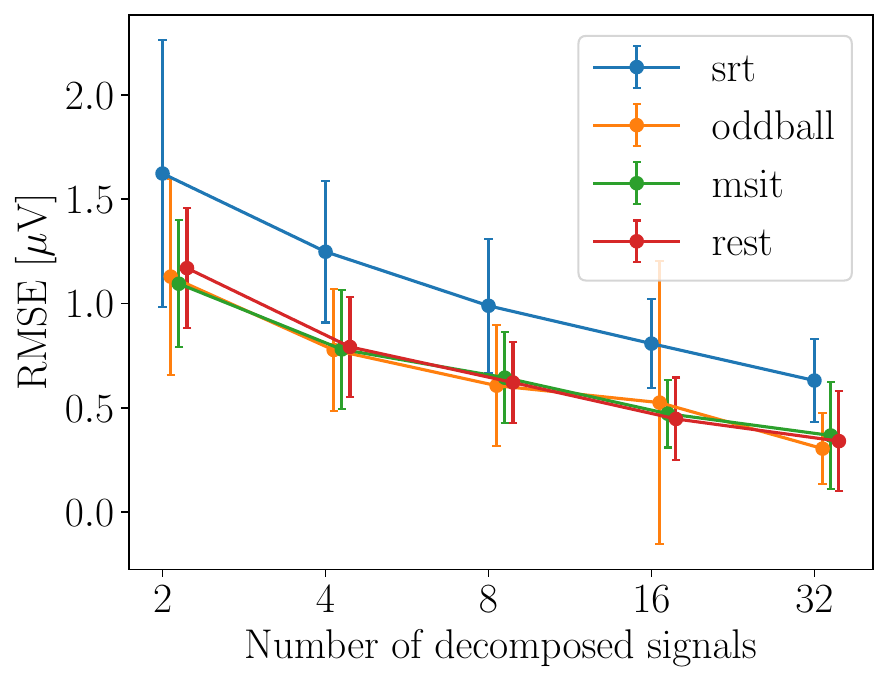}
        \label{fig:reconst:tasks}
    }
    \subfloat[With first $n$]{
        \includegraphics[width=.46\linewidth]{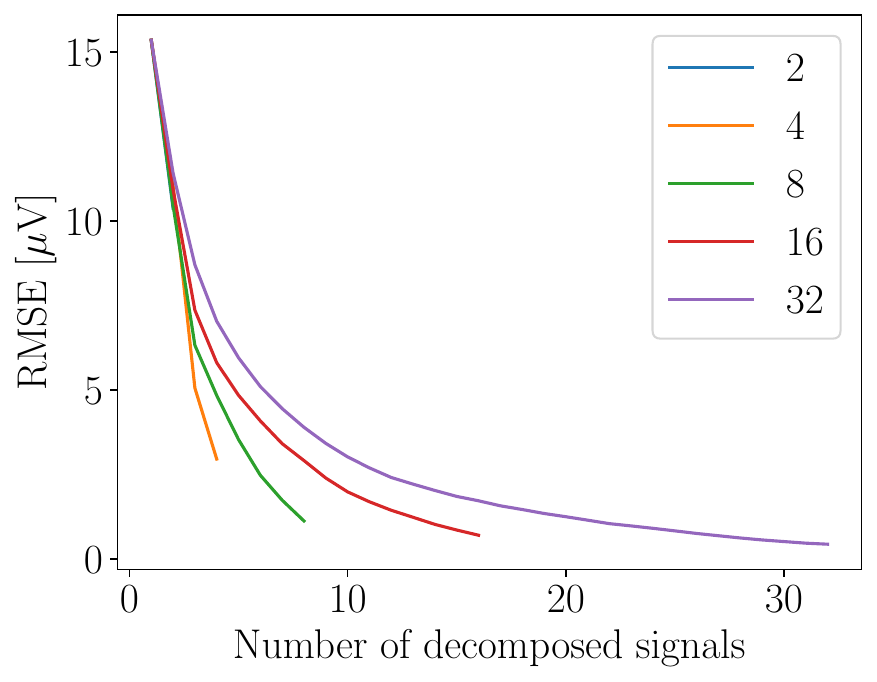}
        \label{fig:reconst:sub_components}
    }
    \\
    \subfloat[Subject/task transfer]{
        \includegraphics[width=.46\linewidth]{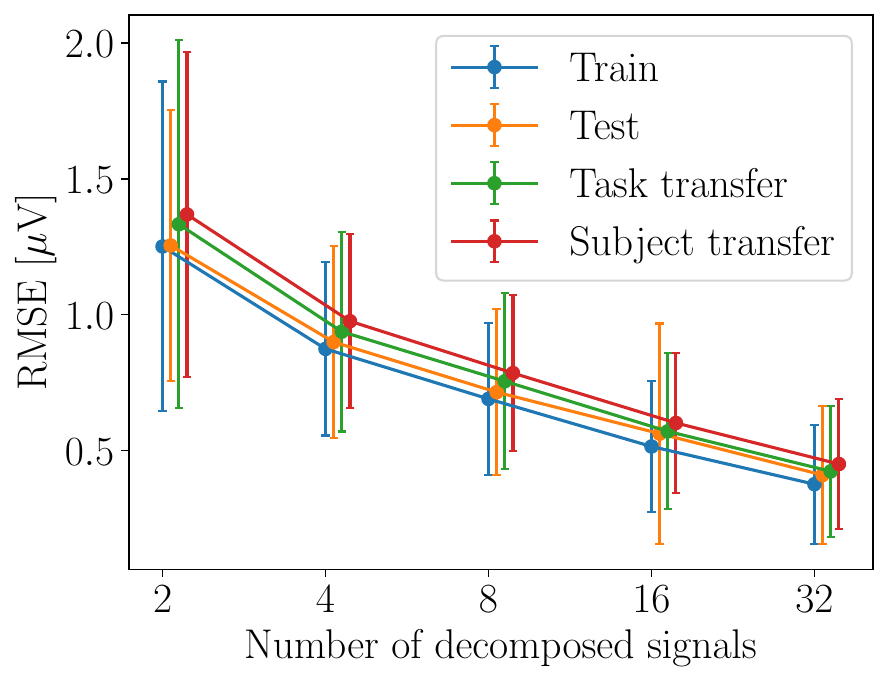}
        \label{fig:reconst:trans}
    }
    \subfloat[Channel transfer]{
        \includegraphics[width=.46\linewidth]{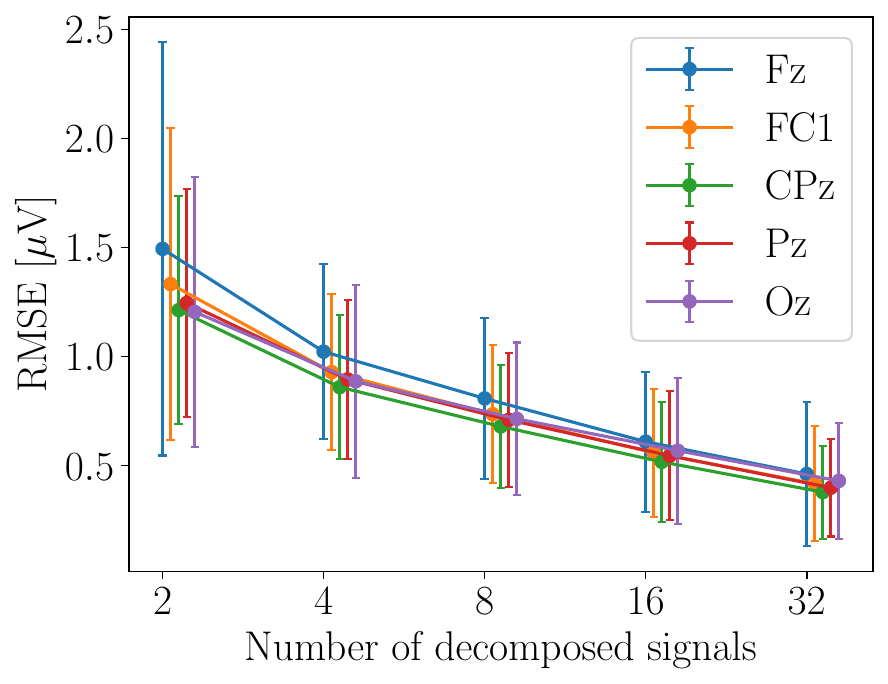}
        \label{fig:reconst:channel}
    }
    \caption{
        Reconstruction accuracy
        (a) for each task,
        (b) with the first $n$ decomposed signals RMSE of which is lowest,
        (c) in the experiment setup for subject and task transfers,
        (d) in the experiment setup for channel transfer.
        The error bars show the standard deviation over session.
    }
    \label{fig:exp:reconst:rmse}
\end{figure}

\begin{figure}[tp]
    \centering
    \subfloat[Sample \#1]{
        \includegraphics[width=.46\linewidth]{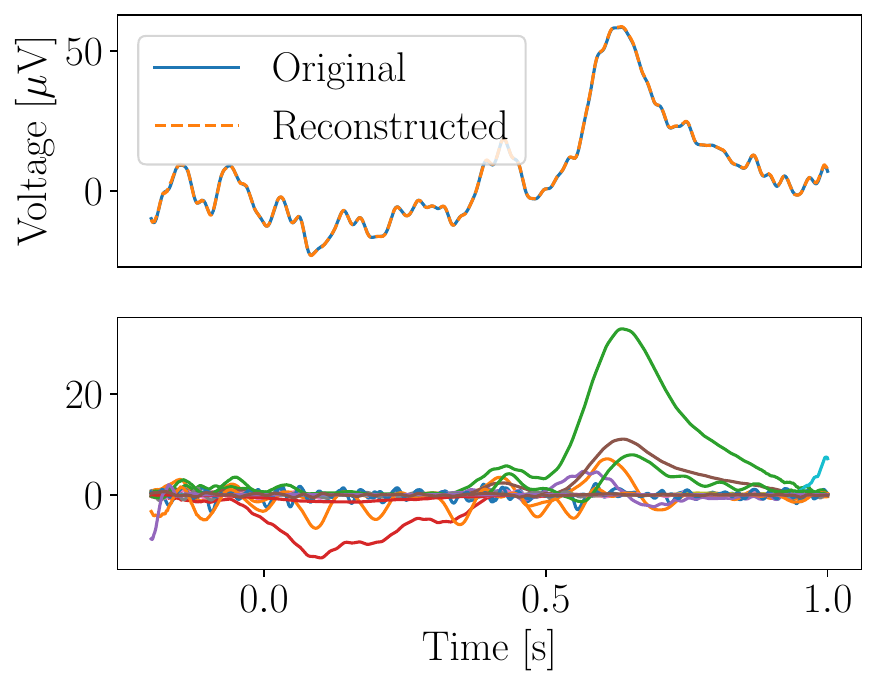}
    \label{fig:reconst:exp1}}
    \subfloat[Sample \#2]{
        \includegraphics[width=.46\linewidth]{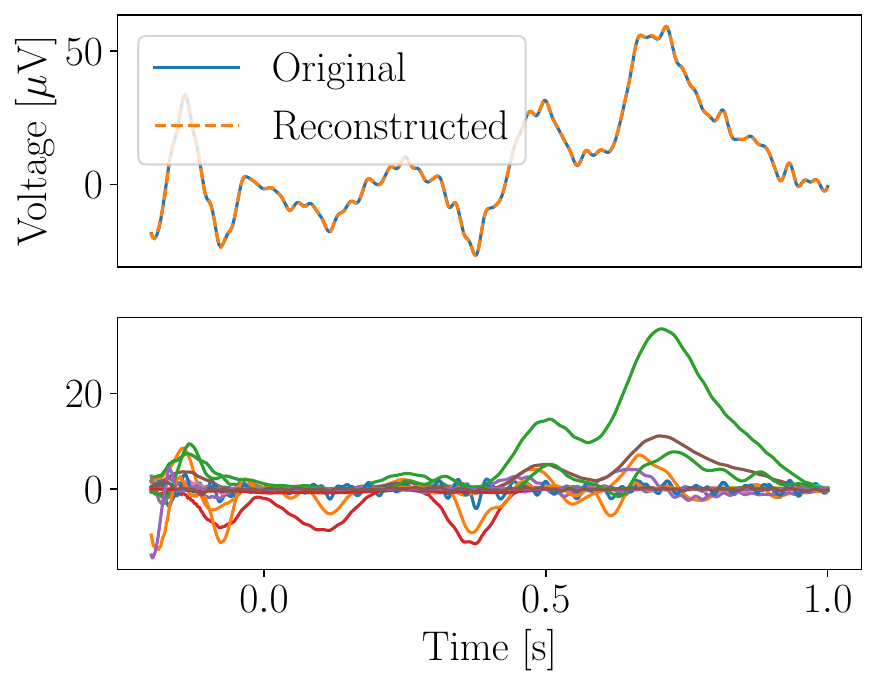}
    \label{fig:reconst:exp2}}
    \caption{
        Examples of original, reconstructed (upper panel), and decomposed (lower panel) signals for different two trials from the testing samples (subject: \#12, task: Oddball, the number of the decomposed signals: 16).
    }
    \label{fig:reconst:exp}
\end{figure}

Figure~\ref{fig:reconst:tasks} shows the reconstruction accuracy for each trial, quantified using the root mean squared error (RMSE), which reflects the average error at each time instance.
For each session, the network was trained on 80\% of the samples and subsequently tested on the remaining 20\%.
For the number of decomposed signals, the results indicated that an increase in the number of decomposed components led to an improvement in RMSE.
Notably, the decomposer configured for 32 signal decompositions achieved an RMSE of approximately 0.5~$\mu$V.
Figure~\ref{fig:reconst:exp} shows examples of the reconstructed signals with RMSEs of 0.16~$\mu$V and 0.19~$\mu$V.
Considering the original signal (approximately $-20$--$50$~$\mu$V), these errors are relatively small, indicating a high level of reconstruction accuracy.
 
\begin{figure}[tp]
    \centering
    \subfloat[Time series]{
        \includegraphics[width=.46\linewidth]{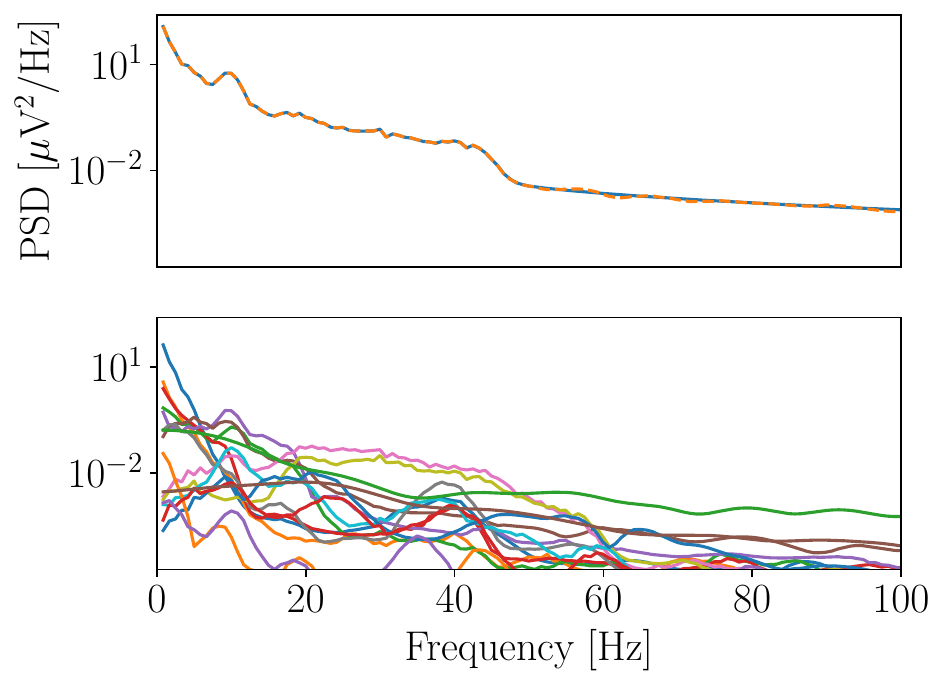}
    \label{fig:reconst:signal}}
    \subfloat[PSD]{
        \includegraphics[width=.46\linewidth]{s12_oddball_Cz_nc016.pdf}
    \label{fig:reconst:spectrum}}
    \caption{
        Averaged original, reconstructed (upper panel), and decomposed (lower panel) signals over trial in the testing samples (subject: \#12, task: Oddball, the number of the decomposed signals: 16).
        (a) Time-series representation of testing samples.
        (b) Power spectrum density (PSD) estimated by a periodogram.
    }
    \label{fig:reconst:ave}
\end{figure}

In the averaged signals over trials, as shown in Figure~\ref{fig:reconst:ave}, the accuracy was remarkably high, as similarly observed in Figure~\ref{fig:reconst:exp}.
In both the time and frequency domains, the lines in the plot would seem to completely overlapped.
The time-domain representation of the decomposed signals exhibited only smooth fluctuations, as these signals were not time-locked, and the high-frequency components were likely diminished through averaging.
However, when examining the power spectrum, computed in the induced response analysis, it became evident that the decomposed signals exhibited distinct frequency properties.
Notably, the frequency bands of some decomposed signals showed overlap.
This suggests that, unlike traditional filter banks, the proposed decomposer does not decompose the signal based on frequency.

Figure~\ref{fig:reconst:sub_components} shows the reconstruction accuracy achieved by using the first $n$ decomposed signals.
In this analysis, the decomposed signals were sorted according to their RMSEs---the error between the original signal and each decomposed signal.
From the figure, it is evident that the decomposers configured with $N = 8, 16, 32$ ultimately attained a lower RMSE level.
This observation suggests that each decomposed signal contributes to the reconstruction process, and the collective effect of these signals on the reconstruction performance varies based on the number of components ($N$) utilized in the decomposition.

Figure~\ref{fig:reconst:trans} shows the results of applying transfer learning.
This figure compares the performance under four conditions:
\begin{itemize}
    \item No transfer (Train and Test):
        This condition is equivalent setup to that for Figure~\ref{fig:reconst:tasks}.
        Both training and testing samples were from the same session.
        The Train condition specifically denotes the use of training samples for testing, offering a baseline for the best possible performance.
    \item Subject transfer:
        The testing samples were from a different subject than the one used for decomposer training, but both the training and testing subjects performed the same task.
        The networks used were identical to those in the no transfer conditions, where 80\% of the samples from a training subject were used to train the decomposer.
        This procedure was repeated for all subject combinations for each task, and the RMSEs were averaged.
    \item Task transfer:
        The training and testing samples were from the same subject, but the subject performed different tasks for the testing and training samples.
        As with the subject transfer condition, the networks used were identical to those in the no transfer condition.
        This procedure was repeated for all task combinations for each subject, and the RMSEs were averaged.
\end{itemize}
As expected, the Train condition yielded the lowest RMSE, followed closely by the Test condition.
Notably, even in scenarios involving subject and task transfers, the RMSEs remained very close to those obtained in the no-transfer condition.
This observation suggests that the trained decomposer was able to generalize effectively across different subjects and tasks, decmonstrating its robustness and versatility in real-world applications.
The generalization capabilities of the decomposer will be further explored in Section~\ref{sec:pre-training}, where we discuss the feasibility benefits of utilzing a pre-trained model of the decomposer.

Figure~\ref{fig:reconst:channel} provides the reconstruction accuracy when applied across different EEG channels.
In this analysis, while maintaining the same 20\% of samples designated for testing (as in previous experiments), the channel that testing signals were recorded was not Cz.
The results indicate that for a decomposer configured with 32 decomposed signals ($N=32$), the RMSE was approximately 1~$\mu$V.
This performance, while slightly inferior compared to the no-transfer condition (where training and testing were done on the same channel), still demonstrates a reasonably high level of accuracy for reconstruction.

\subsection{ERP classification} \label{sec:exp:erp}
In this section, we shifted our focus from assessing the reconstruction accuracy to evaluating the decomposed signals, themselves.
Specifically, we examined their contribution to the classification of ERPs elicited by an oddball task.
The objective of this classification was to accurately determine the category of stimuli presented during the task, based on the EEG signals.

\subsubsection{Dataset} \label{sec:exp:erp:dataset}
The dataset was the same as the one described in Section~\ref{sec:exp:reconst}~\cite{Dzianok2022}.
We exclusively focused on the samples acquired while subjects engaged in the Oddball task.
These samples were labeled as three classes: standard, target, and deviant.
The number of samples for each of the three stimulus classes was equalized for each subject.
This equalization was achieved through a process of random undersampling.
On average, this process yielded a dataset comprising approximately 1,392 samples per subject.

\subsubsection{Architecture} \label{sec:exp:erp:network}
We employed the same decomposers as those used in the experiment detailed in Section~\ref{sec:exp:reconst}.

\subsubsection{Classification procedure} \label{sec:exp:erp:clf}
As previously outlined in Section~\ref{sec:exp:reconst:result}, for the ERP classification experiment, the decomposer was trained using 80\% of the samples from the session while the remaining 20\% were reserved for testing and evaluating classification accuracy.
All the signals decomposed by the decomposer were inputted into a classifier.
We employed a linear support vector machine (SVM)~\cite{Vapnik1998} with a regularization parameter $C$ set to $1.0$.
The primary task of the SVM classifier was to categorize each input EEG signal into one of the three labels.
The classifier itself was trained using the same training samples that were utilized for training the decomposer.
We compared the classification accuracy achieved by using the decomposed signals against that obtained by using the original, undecomposed EEG signals.

\subsubsection{Result} \label{sec:exp:erp:result}
\begin{figure}[tp]
    \centering
    \includegraphics[width=.95\linewidth]{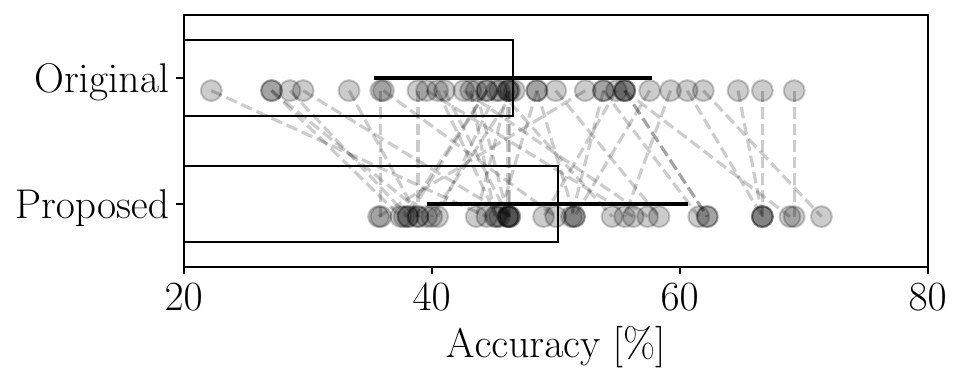}
    \caption{
        Classification accuracy in the oddball task.
        The bars and their error bars represent the average and standard deviation of accuracy across the subjects and the connected grey dots represent the individual accuracy for each subject.
    }
    \label{fig:exp:erp:acc}
\end{figure}

Figure~\ref{fig:exp:erp:acc} presents the classification accuracy results.
The classification that underwent decomposition was significantly higher than those using the original, undecomposed signals
(a paired $t$-test showing $t(39)=-2.821$ and $p=0.007$).

\begin{figure}[tp]
    \centering
    \subfloat[High contribution]{
        \includegraphics[width=.46\linewidth]{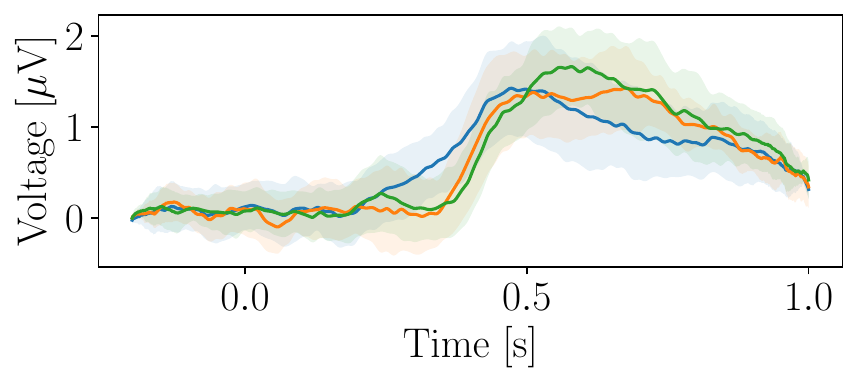}
    \label{fig:exp:erp:sig:hi}}
    \subfloat[Low contribution]{
        \includegraphics[width=.46\linewidth]{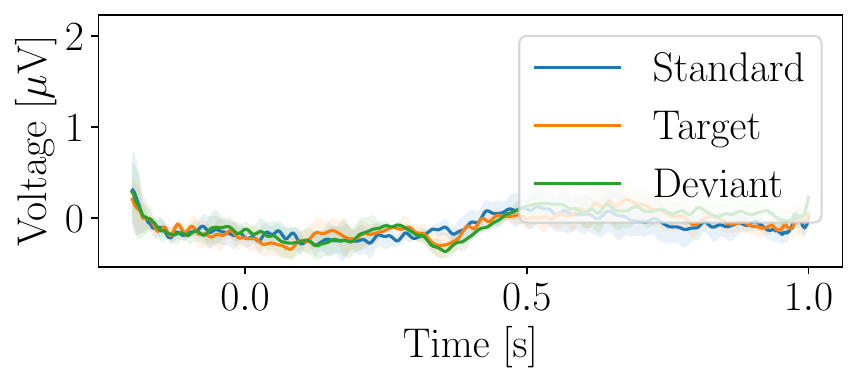}
    \label{fig:exp:erp:sig:lo}}
    \caption{
        Grand-averaged of the decomposed signals, separated by class.
        The shaded area represent the standard deviations of the signals.
        (a) Average of the four decomposed signals with the highest contribution to classification.
        (b) Average of the four decomposed signals with the lowest contribution.
    }
    \label{fig:exp:erp:sig}
\end{figure}

Figure~\ref{fig:exp:erp:sig} shows the grand-average of the decomposed signals that appear to contribute accurate classification.
The contribution of each signal to classification was evaluated by measuring the classification accuracy using only individual decomposed signal, following the same procedure as in Section~\ref{sec:exp:erp:clf}.
Although the decomposer was different for each subject, the decomposed signals with high classification contributions seem to preserve distinctions among the classes.
In contrast, the decomposed signals with low contribution did not appear to have class-specific features.
This suggests that the decomposer effectively separated task-related and non-task-related (possibly noise-related) components, improving the signal-to-noise ratio for certain decomposed signals and contributing to enhanced classification accuracy.

\subsection{Noise reduction} \label{sec:exp:noise}
The effectiveness of the proposed method was further assessed in terms of its performance in noise reduction.
Instead of direct evaluation of noise reduction performance, we evaluated its impact indirectly by measuring classification accuracy in a motor-imagery (MI)-based BCI dataset.

\subsubsection{Dataset} \label{sec:exp:noise:dataset}
For our noise reduction performance assessment, we utilized an open dataset, specifically designed for MI-based BCIs, as detailed in~\cite{Ofner2017}.
This dataset comprises samples from 15 subjects, each engaging in seven distinct MI tasks.
The tasks included in the datasets are $C_{1}$ (elbow flexion), $C_{2}$ (elbow extension), $C_{3}$ (supination), $C_{4}$ (pronation), $C_{5}$ (hand close), $C_{6}$ (hand open), and $C_{7}$ (rest).
The dataset also encompasses three electrooculography (EOG) channels, which record the electrical activity of the muscles around the eyes.
The electrodes for three channels were placed around the left, right, and top areas of the eyes.
EOG data is useful for identifying artifacts related to eye movements, which are common sources of noise in EEG signals.

The original EEG were recorded using 61 electrodes at a sampling rate of 512~Hz.
However, given that our study concentrates on single-channel decomposition, we exclusively utilized data from channel F1.
Channel F1 is situated in the frontal area of the head, a region frequently susceptible to noise events, such as eye blinks and saccades.
For each subject, the dataset comprised 120 samples.

\subsubsection{Architecture} \label{sec:exp:noise:network}
The decomposer was configured based on the network architecture detailed in Section~\ref{sec:mm:noise}.
Both the signal-estimator and noise-estimator within the network designed to provide 16 decomposed signals each.
The detectors in both estimators consisted of four layers, with each layer comprising a convolutional layer (having a kernel size of 33, which equates to approximately 0.13~s, and single channel), followed by a ReLU activation function.
The convolutional layer in the atom network was set with a kernel size of 512 corresponding to roughly 2~s.

The optimization of the networks for both the signal- and noise-estimators was carried out using supervised loss functions, as specified in (\ref{eq:signal}) and (\ref{eq:noise}).
For this process, an Adam optimizer (a learning rate of 0.0001, $\beta_{1}$ of 0,5, $\beta_{2}$ of 0.999, and a weight decay of $10^{-5}$) was employed.
The optimization was spanned over 2,000 iterations (epochs) with a batch size of 100.
In the case of the noise-estimator, we used the loss function in (\ref{eq:noise}) for the initial 1,000 iterations, followed by (\ref{eq:noise2}) for the subsequent 1,000 iterations.

\subsubsection{Classification procedures} \label{sec:exp:noise:procedure}
For the classification task, the available samples were categorized into three separate sets, serving distinct purposes in the experimental procedure.
The first set is samples for decomposer training, denoted by $\mathcal{R}_{1}$.
These samples were selected from the classes $C_{1}$, $C_{2}$, $C_{4}$, $C_{5}$, and $C_{6}$.
They were specifically used for training the decomposer.
The second and third sets are samples for classifier training, $\mathcal{R}_{2}$, and for testing, $\mathcal{R}_{3}$.
These sets comprised samples from classes $C_{3}$ and $C_{7}$.
They were divided into $\mathcal{R}_{2}$ and $\mathcal{R}_{3}$ using a leave-one-out cross validation (CV) method.

We trained the decomposer using the samples from set $\mathcal{R}_{1}$.
The noise-estimators were optimized based on labels indicating the presence or absence of noise events at each time index.
These labels were derived from the EOG signals.
From the three EOG channels available in the dataset, we derived three bipolar EOG signals.
Time indexes where the amplitude of these bipolar signals exceeded the threshold of $\pm50$~$\mu$V were marked as instances of noise-event-present.

The decomposer, trained with these noise-event labels, was then able to differentiate and produce two distinct outputs for each sample:
one representing the noise-related signal and the other presenting the brain-related signal.
The next step of the process involved the classification of these brain-related signals, i.e., the outputs of the signal-estimator.
For this purpose, we employed an SVM classifier equipped with a radial basis function kernel.
The SVM classifier was configured with a kernel coefficient $\gamma$ set to ``scale'' and a regularization parameter $C$ set to 1.
The classifier's task was to discriminate an output of the signal-estimator into either class $C_{3}$ or $C_{7}$.

In addition to the proposed method (shown as {\it Proposed}), performance was also evaluated under two alternative conditions for comparison: {\it Raw}, {\it Reject}, {\it EOG-Reg}, and {\it SSA-ICA}.
In the Raw condition, the signal underwent no decomposition or noise reduction.
The observed signals in $\mathcal{R}_{2}$ were directly input into the classifier without any processing.
In the Reject condition, a straightforward noise rejection approach was implemented.
Any samples with time indexes labelled as having noise events were completely excluded.
In the EOG-Reg condition, we applied a method for reducing EOG artifacts through regression~\cite{Croft2000}, as implemented in MNE-Python~\cite{Gramfort2013}.
In the SSA-ICA condition, we used a frequency-based single-channel decomposition method that combines singular spectrum analysis (SSA) and ICA~\cite{Maddirala2018}.
We first decomposed the single-channel signal using SSA into eight signals and then applied ICA to the eight decomposed signals.
Independent components with a correlation coefficient greater than $0.7$ or less than $-0.7$ with any EOG bipolar signals were removed, and the signal was reconstructed using the inverse transforms of ICA and SSA.
It should be noted that this comparison is not entirely fair, as the proposed method requires training data, whereas both the EOG-Reg and SSA-ICA methods only require the observed signal for decomposition.

\subsubsection{Result}
\begin{figure}[tp]
    \centering
    \subfloat[Sample \#1]{
        \includegraphics[width=.46\linewidth]{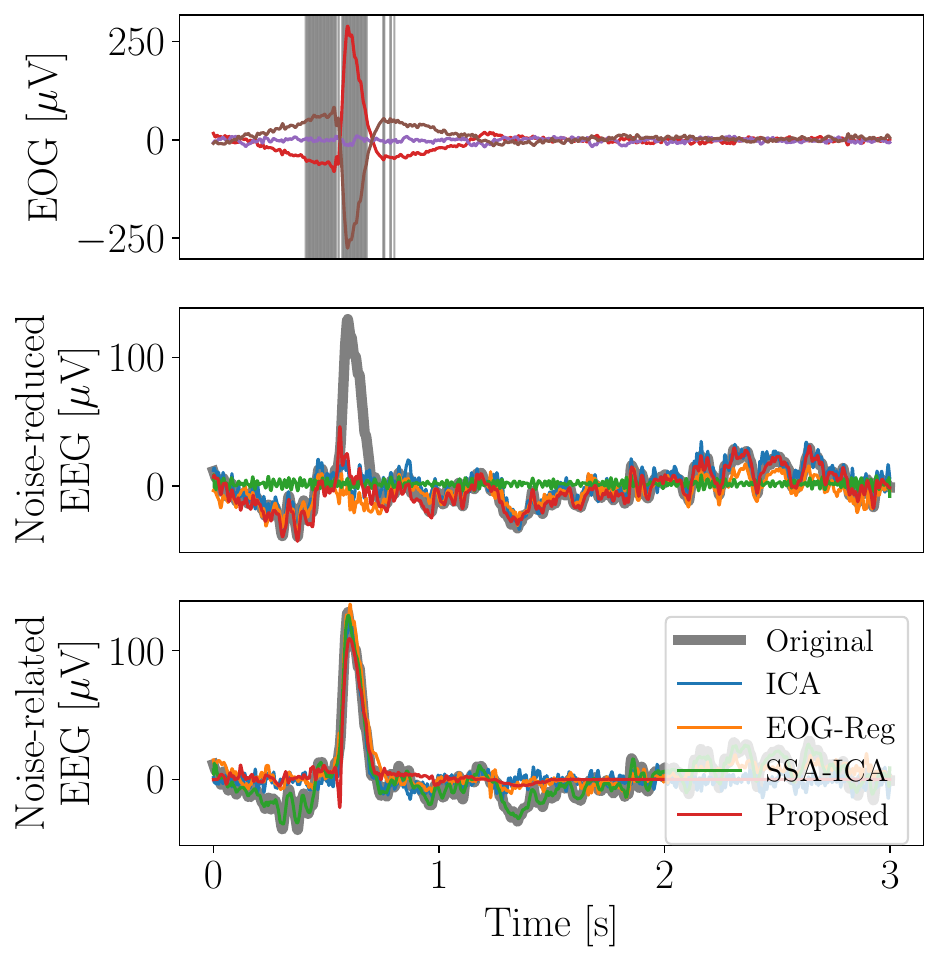}%
    \label{fig:noise:signals:a1}}
    \subfloat[Sample \#2]{
        \includegraphics[width=.46\linewidth]{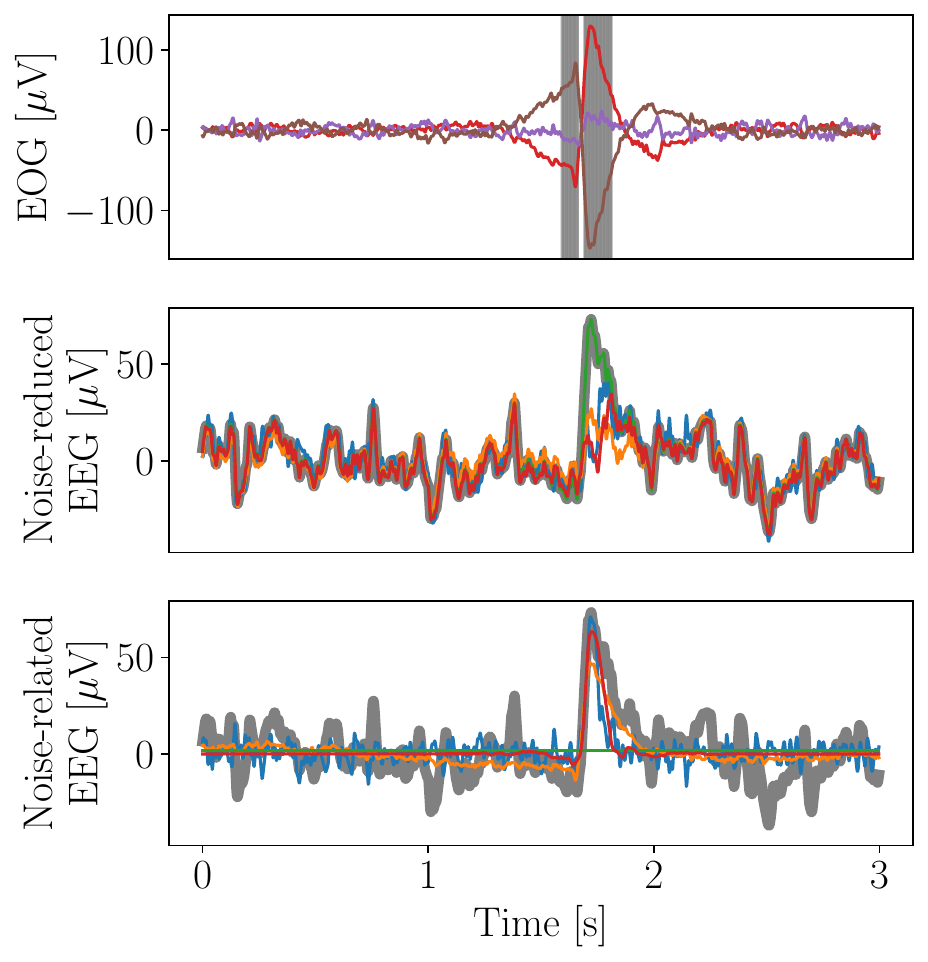}%
    \label{fig:noise:signals:a2}}
    \\
    \subfloat[Sample \#3]{
        \includegraphics[width=.46\linewidth]{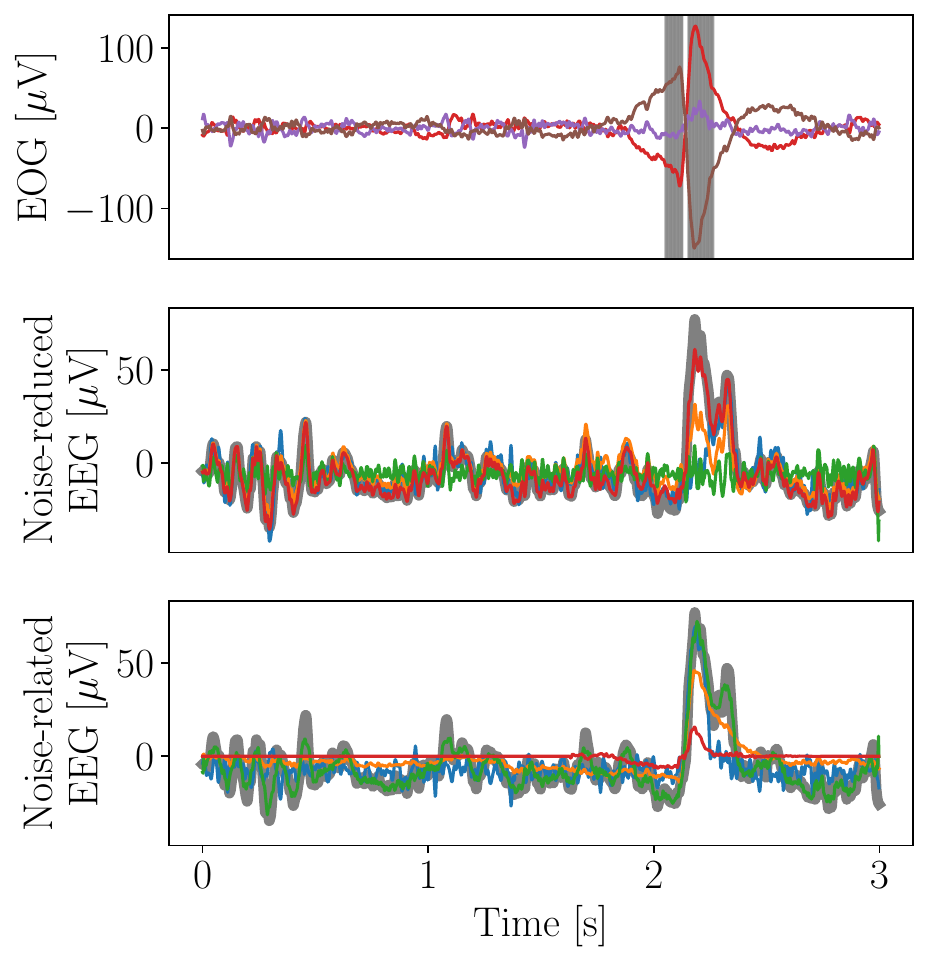}%
    \label{fig:noise:signals:b}}
    \subfloat[Sample \#4]{
        \includegraphics[width=.46\linewidth]{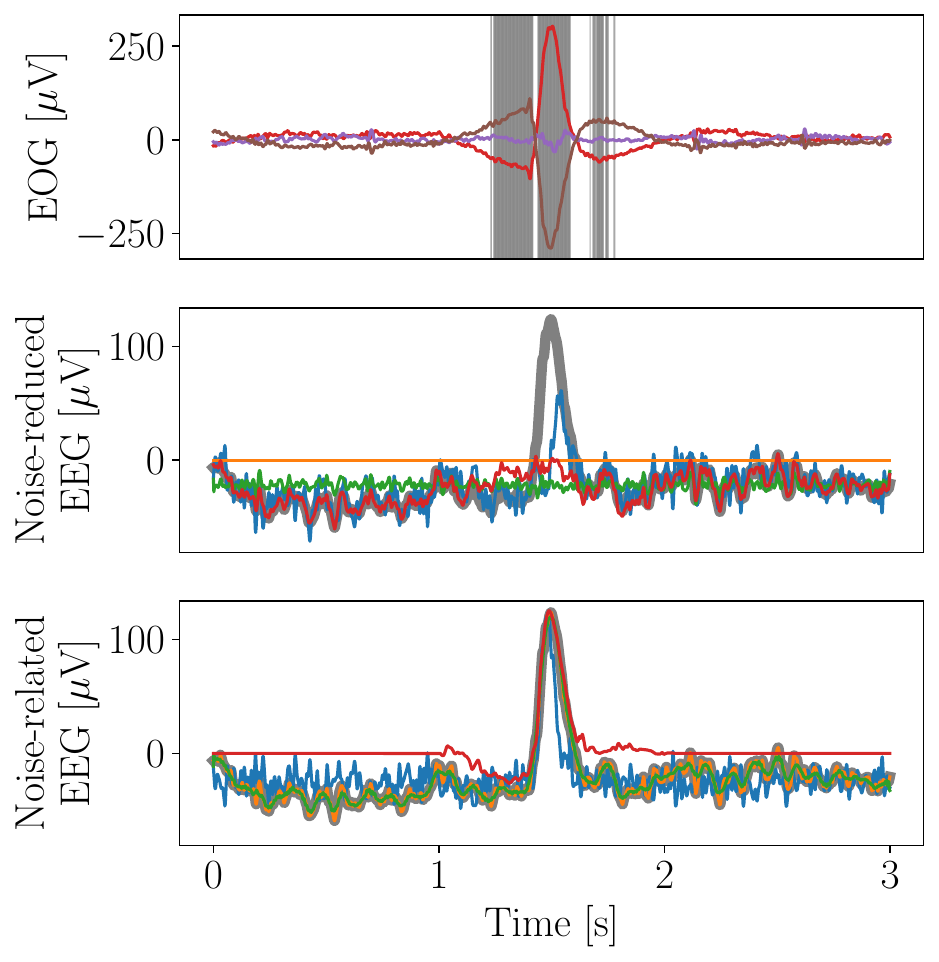}%
    \label{fig:noise:signals:c}}
    \caption{
        Examples of signal samples from the sample set $\mathcal{R}_{2}$.
        The top panel displays the bipolar signals of the three EOG reference signals (TL: top and left channels, LR: left and right, RT: right and top).
        The middle panel displays the original EEG signal and noise-reduced signals.
        The bottom panel displays the original EEG signal and noise-related signals.
    }
    \label{fig:noise:signals}
\end{figure}

Figure~\ref{fig:noise:signals} presents several signal examples.
The EOG signals shown in the top panel were used for labeling each time index as noise-event-absent or noise-event-present.
The grey-shaded areas represent time indexes marked as noise-event-present, where large amplitude fluctuations in the EOG signals suggest the present of noisy events like eye blinks or eye movements.
The middle panel compares the original EEG signals (channel F1) with their noise-reduced counterparts.
In addition to EOG-Reg, SSA-ICA, and the proposed method, the noise reduction was applied by ICA which utilized signals from all 64 EEG channels.
For the noise-reduced signals via ICA, independent components with a correlation coefficient greater than $0.7$ or less than $-0.7$ with any EOG bipolar signals were removed from the reconstruction.

Figures~\ref{fig:noise:signals:a1} and~\ref{fig:noise:signals:a2} indicate that both EOG-Reg, ICA, and the proposed method were all effective in reducing large fluctuations presumed to be caused by noise events.
In contrast, SSA-ICA altered the signal values even during periods without noise events, suggesting that frequency-based decomposition may not be suitable for reducing EOG artifacts.
As shown in Figure~\ref{fig:noise:signals:b} for the proposed method and Figure~\ref{fig:noise:signals:c} for ICA and EOG-Reg, there were instances where both methods did not completely eliminate these fluctuations.
This observation suggests that while these noise reduction techniques are generally effective, they are not foolproof in all scenarios.

\begin{figure}[tp]
    \centering
    \includegraphics[width=.95\linewidth]{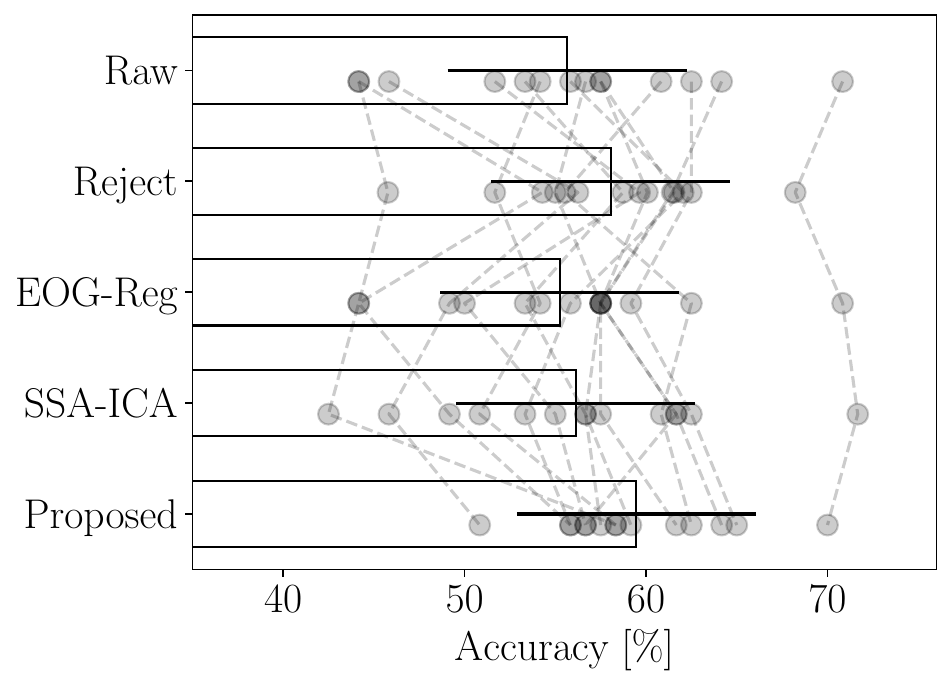}
    \caption{
        Classification accuracy for the MI task.
        The bars and their error bars represent the average and standard deviation of accuracy across the subjects and the connected grey dots show the accuracy of individual subjects.
    }
    \label{fig:noise:acc}
\end{figure}

Figure~\ref{fig:noise:acc} displays the classification accuracy rates for each condition in the MI task.
The proposed method demonstrated significantly higher accuracy rates compared to the Raw, EOG-Reg, and SSA-ICA conditions, with $p$-values of 0.007, 0.005, and 0.022, respectively.
Both the EOG-Reg and SSA-ICA conditions appear to reduce not only noise but also other signal components, as shown in Figure~\ref{fig:noise:signals:c}, which may lead to the decline in accuracy.
Furthermore, the accuracy rate by the proposed method were on par with those of the Reject condition, evidenced by a non-significant difference with a $p$-value of 0.298.
This suggests that the effectiveness of the proposed method in reducing noise-related components and its subsequent contribution to a robust classification.

\subsection{SSVEP classification} \label{sec:exp:ssvep}
In this section, we evaluated the performance of the proposed decomposer, specifically designed with an architecture where a single atom is shared across all detectors.
This unique design aligns with the signal model associated with SSVEP-BCI.

\subsubsection{Dataset} \label{sec:exp:ssvep:dataset}
%
Two open datasets~\cite{Wang2016a,Liu2020} designed for SSVEP-BCI research were used in this experiment.
These datasets collectively included EEG recordings from 35 and 70 subjects, respectively.
Subjects were displayed 40 stimuli, each flickering at different frequencies and arranged in a matrix on a monitor.
The task is to identify the specific stimulus a subject was gazing at, based on solely on the EEG signal.
The original signals in both datasets were recorded using 64 channels at a sampling rate of 250~Hz.
However, for the purpose of this SSVEP-BCI task, only the data from channel Oz were used.
Signals were epoched in the time range of $0$--$3$~s post stimulus onset for~\cite{Wang2016a} and $0$--$2.5$~s for~\cite{Liu2020}.
Additionally, the epoch signals underwent bandpass filtering using a 4th order Butterworth filter with a frequency range of 5 to 100~Hz.
For each subject, each class had 6 samples, leading to a total of 240 samples for~\cite{Wang2016a}.
For~\cite{Liu2020}, there were 4 samples per class, resulting in 160 samples in total.

\subsubsection{Architecture} \label{sec:exp:ssvep:network}
The decomposer was configured based on the network architecture described in Section~\ref{sec:mm:ssvep}.
The decomposer was designed to output a number of decomposed signals corresponding to the number of stimuli, which in this case was 40.
Each detector had two layers.
Each layer consisted of a convolutional layer with a kernel size of 501 (equivalent to approximately 2~s) and two channels.
The final layer in the detector sequence employed a ReLU activation function.
The detectors shared a single atom.
The length of this atom was set to 125, corresponding to 0.5~s.
The supervised loss, as defined in (\ref{eq:ssvep_loss}), was used for the optimization of the decomposer.
For the optimization, an Adam optimizer was employed with the following settings: a learning rate of 0.0001, $\beta_{1}$ of 0.5, $\beta_{2}$ of 0.999, and a weight decay of $10^{-5}$.
The optimization process spanned over 200 iterations (epochs), and was conducted with a batch size of 40.

\subsubsection{Classification Procedure} \label{sec:exp:ssvep:procedure}
For the SSVEP classification, the procedure involved splitting the samples into two equal parts:
one set for training and the other for testing.
The training samples were first utilized for optimizing the decomposer.
Following the optimization of the decomposer, the same training samples were then used to train a classifier.
For classification, the powers of the decomposed signals, derived from the decomposer, were used as input features.
The classifier was implemented using an SVM with a kernel coefficient $\gamma$ was set to ``scale'' and a regularization parameter $C$ of 1.
In addition to using the decomposed signals, the experiment also included a comparison where the power spectrum density (PSD) of the original EEG signal was used as input to the classifier.

\subsubsection{Result} \label{sec:exp:ssvep:result}
\begin{figure}[t]
    \centering
    \includegraphics[width=.95\linewidth]{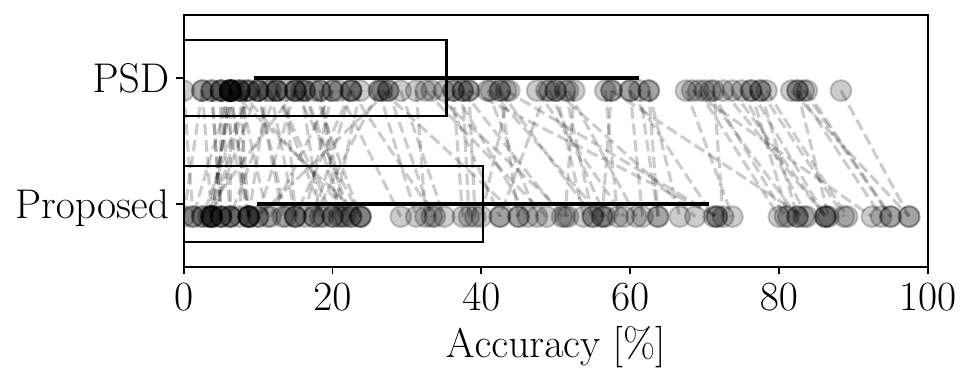}
    \caption{
        Classification accuracy for the SSVEP classification task.
        The bars and their error bars represent the average and standard deviation of accuracy across subjects, while the connected grey dots show the accuracy of individual subjects.
    }
    \label{fig:ssvep:acc}
\end{figure}

Figure~\ref{fig:ssvep:acc} shows the classification accuracy rates for the SSVEP-BCI task, encompassing results from 105 subjects.
The results revealed that the average rate was 35.3\% for the method using PSD features and 40.23\% for the proposed method.
A paired $t$-test was conducted to validate this improvement statistically, yielding $t(104)=-5.995$ and $p < 0.001$,
which clearly demonstrates the significance of the accuracy enhancement brought by the proposed method.
Notably, the improvement was particularly pronounced among subjects who already had relatively high accuracy with the PSD method.
This observation suggests that a robust amplitude of SSVEP signals might be a critical factor for the effective optimization of the decomposer.

\begin{figure}[t]
    \centering
    \subfloat[Detector output]{
        \includegraphics[width=.46\linewidth]{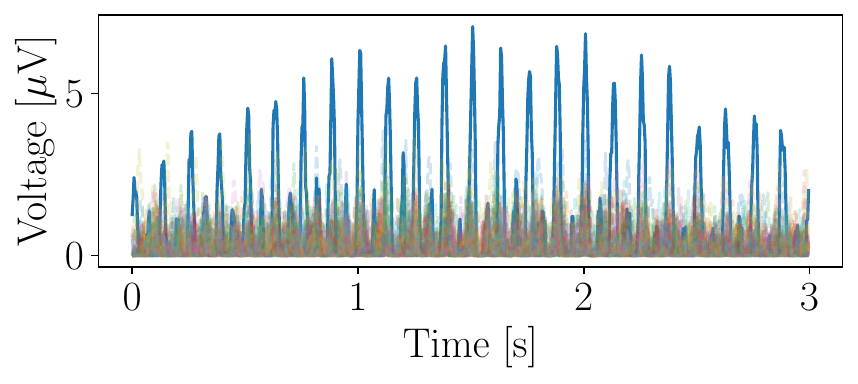}%
        \includegraphics[width=.46\linewidth]{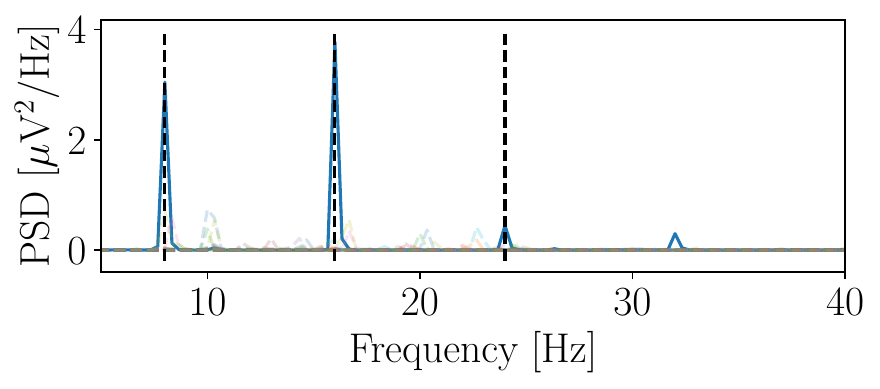}%
    \label{fig:ssvep:det}}
    \\
    \subfloat[Atom]{
        \includegraphics[width=.46\linewidth]{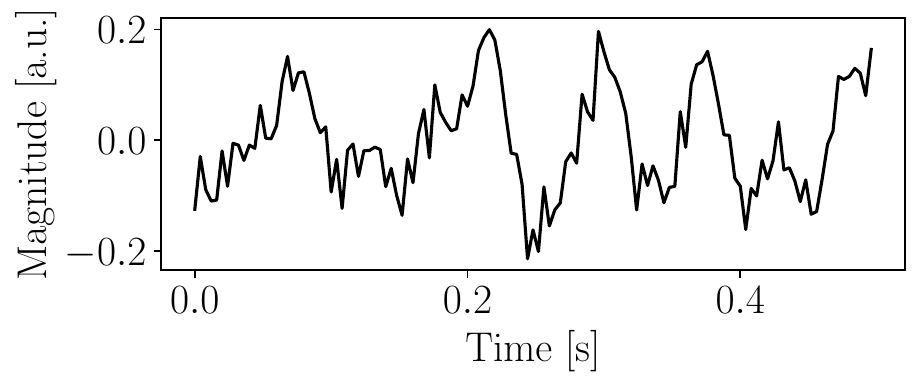}%
    \label{fig:ssvep:atom}}
    \subfloat[Decomposed]{
        \includegraphics[width=.46\linewidth]{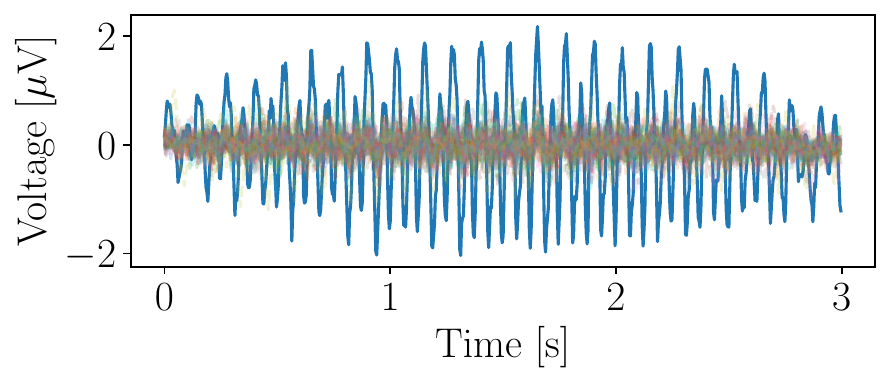}%
    \label{fig:ssvep:decomp}}
    \\
    \subfloat[Reconstructed]{
        \includegraphics[width=.46\linewidth]{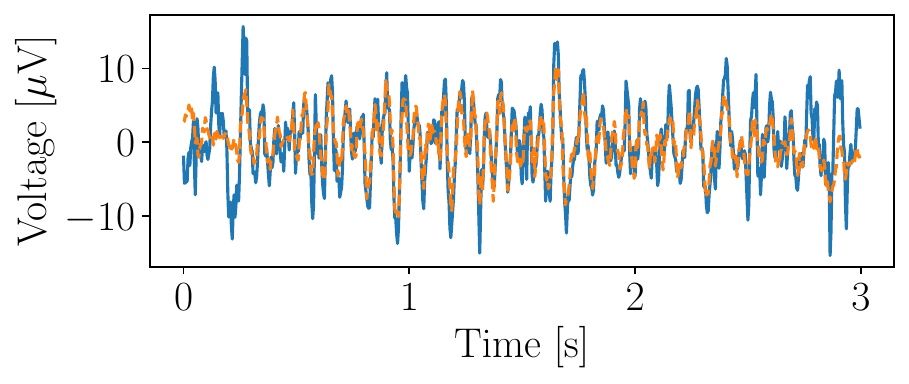}%
        \includegraphics[width=.46\linewidth]{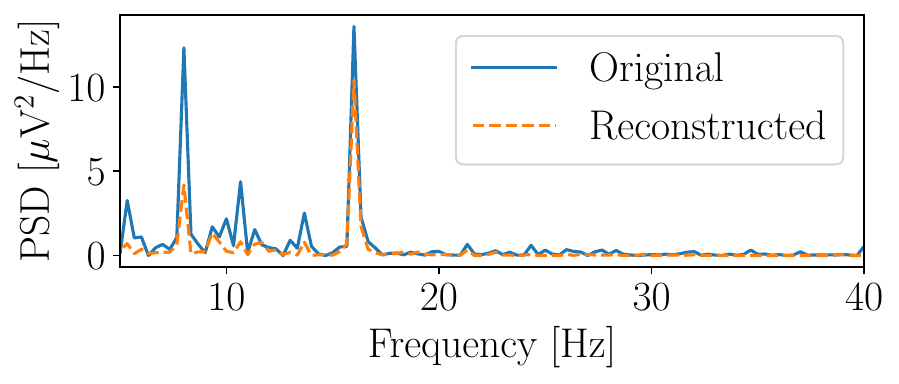}%
    \label{fig:ssvep:reconst}}
    \caption{
        Example of a signal sample in the SSVEP classification.
        (a) Detector outputs (time and frequency domains).
        The output corresponding to the class of the signal sample (target) is highlighted with a solid line, while the outputs for others classes are shown with dashed lines.
        In the frequency domain, the black dashed lines indicate the flickering frequency of the target class stimulus and its harmonics.
        (b) Atom shared by all decomposed signals.
        (c) Decomposed signals.
        The signal corresponding to the class label of the sample is drawn with a solid line.
        (d) Original and reconstructed signals.
    }
    \label{fig:ssvep:signal}
\end{figure}

Figure~\ref{fig:ssvep:signal} provides a detailed view of a single-trial signal sample.
Figure~\ref{fig:ssvep:det} show outputs of the detectors, with each detector tailored for a specific class among the 40 available.
This PSD suggests that the detector captured the target stimulus's flickering frequency and its harmonics.
The atom, which is shared across all decomposed signals, is depicted in Figure~\ref{fig:ssvep:atom}.
It represents an estimation of the VEPs induced by a single flash, composing the flickering stimulus.
As illustrated in Figure~\ref{fig:ssvep:decomp}, The decomposed signal identified by the detector corresponding to the target stimulus class exhibited greater power compared to the other decomposed signals.
The reconstructed signal shown in Figure~\ref{fig:ssvep:reconst} does not precisely replicate the original, reflecting the decomposer's focus on class separation rather than pure reconstruction fidelity.
The spectra of both the original and reconstructed signals show that the reconstruction process may enhance certain frequency components characteristic of SSVEP.

\subsection{ERP decomposition} \label{sec:exp:err}
In this section, we explored the application of the proposed decomposer in extracting error-related potentials (ErrPs).
ErrPs are specific types of ERPs that emerge in response to negative feedback~\cite{Nieuwenhuis2004}.
They have been recognized for their usefulness in detecting errors within BCI systems~\cite{Parra2003}.

While we previously assessed our decomposer's capability with ERPs in Section~\ref{sec:exp:erp}, ErrPs are characterized by their time-invariant nature across trials.
This specific attribute of ErrPs, where there is no expected time shift in response to stimuli or events, makes the specialized decomposer architecture described in Section~\ref{sec:mm:time-locked}.

\subsubsection{Dataset} \label{sec:exp:err:dataset}
We used an open dataset designed for error detection during a P300 spelling task~\cite{Margaux2012}.
This dataset included samples from 15 subjects.
The dataset posed a classification challenge where EEG signals are classified into either ``Error'' or ``Correct'' classes.
These classes represented whether the feedback from the BCI corresponds to the target letter the subject intended to spell.
The EEG data were originally recorded using 56 channels at a 200~Hz sampling rate.
However, in line with the focus of our study on single-channel decomposition, we used data from channel Cz.

As preprocessing, the continuous signals were filtered using a 5th order Butterworth filter within a frequency range of 1 to 40~Hz.
The signals were segmented into epochs spanning from $-0.1$--$1.3$~s relative to the onset of the BCI system feedback.
The epoch samples were baseline-corrected using the average voltage calculated over a period from $-0.2$ to $0$~s.
Any epoch samples that exhibited voltage excursions beyond $\pm 80$~$\mu$V were excluded from further analysis.
After these preprocessing steps, a total of 3,254 samples remained for analysis.

For this experiment, our aim was to identify components common across all subjects.
To achieve this, a single decomposer was designed for all the subjects' samples, akin to the concept of grand-averaging, which typically involves averaging signals across multiple subjects.
Consequently, samples from all subjects were concatenated into a unified dataset.

\subsubsection{Architecture} \label{sec:exp:err:network}
The decomposer was structured based on the framework introduced in Section~\ref{sec:mm:time-locked}.
The decomposer was set up to decompose an input signals into eight distinct signals.
These eight detector-atom pairs shared a uniform architecture.
The detector of each network consisted of three convolutional layers, each with a kernel size of 25, translating to 125~ms, followed by a linear layer and a ReLU activation function.
The atom network was designed with a kernel size of 280 (1.4~s), equivalent to the signal length of the input.
The networks were optimized using the supervised loss function as detailed in (\ref{eq:loss:supervised}).
Half of the networks were dedicated to the Error class, with optimization aimed at accurately reproducing signals characteristic of error responses.
The remaining networks were focused on the Correct class.
This architecture and loss function aimed to effectively distinguish between common and unique signatures associated with both Error and Correct classes.
The network underwent optimization process, spanning 10,000 iterations (epochs).
The batch size was set at 100.

\subsubsection{Result} \label{sec:exp:err:result}
\begin{figure}[t]
    \centering
    \subfloat[Original]{
        \includegraphics[width=.46\linewidth]{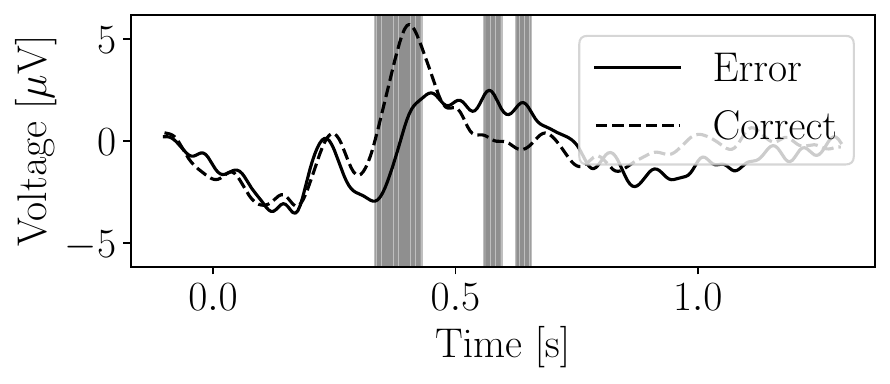}%
    \label{fig:err:original}}
    \subfloat[Reconstructed]{
        \includegraphics[width=.46\linewidth]{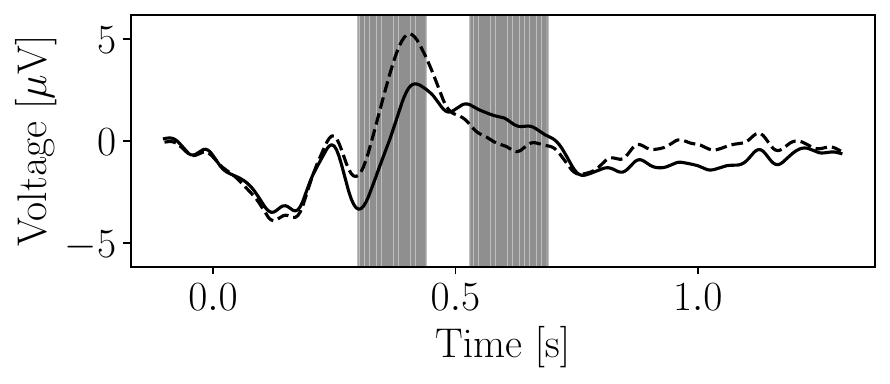}%
    \label{fig:err:reconst}}
    \caption{
        Grand-averaged signals of the original and reconstructed ones for each class in the ErrP dataset.
        The grey shaded areas highlight the significant differences between the classes, based on a $t$-test ($p < 0.001$).
    }
    \label{fig:err}
\end{figure}

The performance evaluation was conducted using a setting where 50\% of the samples were randomly selected for training the decomposer, with the remaining 50\% used for testing.
Although, our primary objective is to identify EEG signatures between classes, necessitating the use of all samples for training, this setup of splitting the training and testing samples was essential to avoid overfitting, which could result in spurious EEG signatures.
The waveforms presented in the results were decomposed and reconstructed from the training samples.
The RMSEs for the training and testing samples were 8.41~$\mu$V and 8.55~$\mu$V, respectively.
This RMSE was notably higher compared to the results in Section~\ref{sec:exp:reconst}.
This difference in error magnitude can be attributed to the challenges of reconstructing EEG signals from 15 subjects using only eight atoms without any time shift.
Figure~\ref{fig:err} shows the grand-averaged original and reconstructed signals for each class.
Even though these signals are grand-averaged, which is expected to reduce reconstruction errors caused by noise, we still observe some slight errors through visual comparison, suggesting that the reconstruction accuracy was not particularly high.

\begin{figure}[t]
    \centering
    \subfloat[]{
        \includegraphics[width=.46\linewidth]{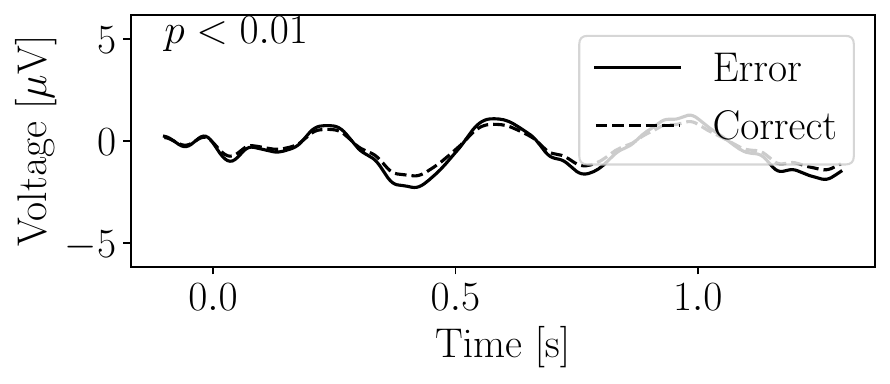}%
    \label{fig:err:dec:01}}
    \subfloat[]{
        \includegraphics[width=.46\linewidth]{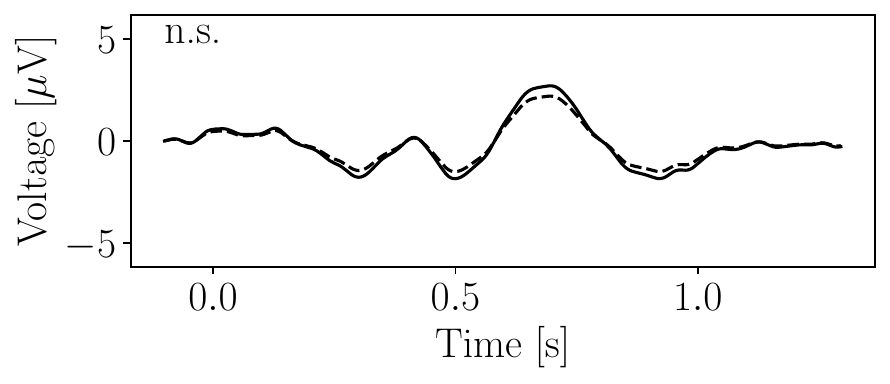}%
    \label{fig:err:dec:02}}
    \\
    \subfloat[]{
        \includegraphics[width=.46\linewidth]{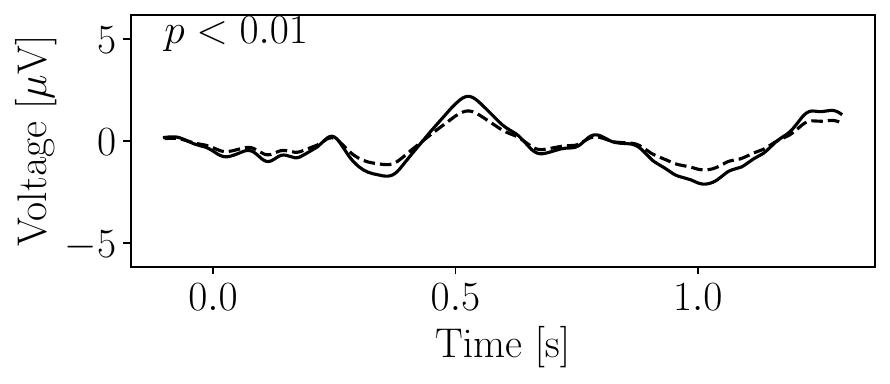}%
    \label{fig:err:dec:03}}
    \subfloat[]{
        \includegraphics[width=.46\linewidth]{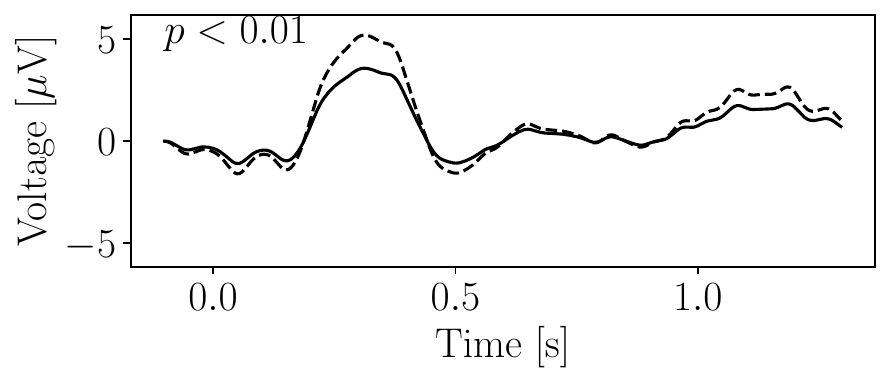}%
    \label{fig:err:dec:05}}
    \caption{
        Grand-averaged decomposed signals, selected four out of eight, for each class in the ErrP dataset.
        The $p$-values indicate significant differences in detector outputs between classes, as determined by a $t$-test.
    }
    \label{fig:err:decomposed}
\end{figure}

\begin{figure}[t]
    \centering
    \subfloat[Different]{
        \includegraphics[width=.46\linewidth]{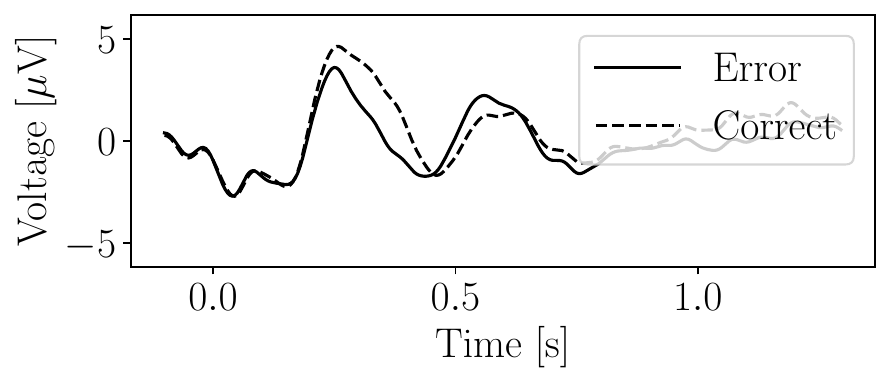}%
    \label{fig:err:reconst:diff}}
    \subfloat[Common]{
        \includegraphics[width=.46\linewidth]{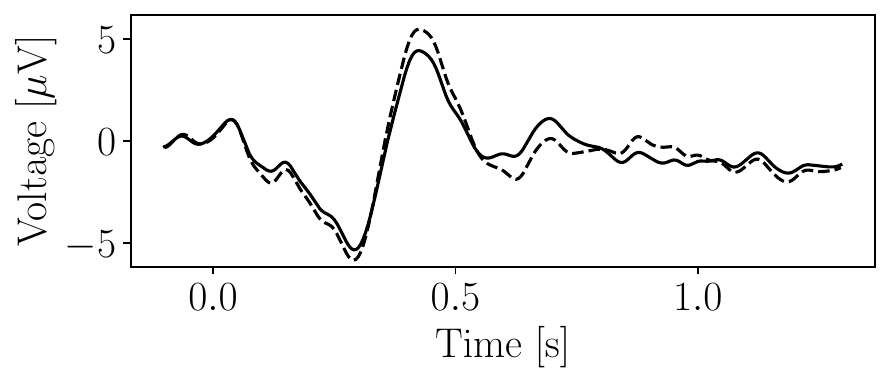}%
    \label{fig:err:reconst:common}}
    \caption{
        Grand-averaged signals reconstructed using the decomposed signals showing significant differences between the class (a) and not showing (b).
    }
    \label{fig:err:reconst2}
\end{figure}

Figure~\ref{fig:err:decomposed} presents the grand average of the decomposed signals, offering insights into the effectiveness of the decomposer in capturing class-specific components.
The $p$-values indicated in the figures were calculated through a $t$-test.
This statistical test was applied to evaluate whether the scalar outputs by the detectors (i.e., $d_{n}(\bm{x})$ in (\ref{eq:model_erp2})) differed significantly between the Error and Correct classes.
A significant difference in these values implies that the decomposers have successfully captured components characteristic of each class.
Three decomposed signals show significant differences, while the others did not.
Moreover, Figure~\ref{fig:err:reconst2} illustrates two reconstructed signals.
One (Different) was reconstructed using output signals from the decomposers that demonstrated significant differences between classes.
These can be interpreted as the ERPs specifically modulated by  the class distinction.
Another was reconstructed using outputs from the decomposers where no significant class-based difference was observed.
These represent components that are common across both classes.

This demonstration provides an intriguing perspective on the nature of ErrPs, traditionally understood in neuroscience as negative voltage fluctuations in response to error observation.
The decomposed signals, as illustrated in Figures~\ref{fig:err:decomposed} and~\ref{fig:err:reconst:diff}, suggest that a neural response typically recognized as an ErrP might actually be the result of overlapping multiple neural components.
The waveforms of the decomposed signals suggest that the observation of an error might lead to a reduction in the amplitude of a positive fluctuation, rather than solely generating a negative fluctuation.
Although further investigation is required to understand the specific conditions under which ERPs are activated or suppressed, our decomposer could be a effective tool for  disentangling these components.

\subsection{Pre-trained decomposer and cross-dataset validation} \label{sec:pre-training}
The results of the transfer learning shown in Figure~\ref{fig:exp:reconst:rmse} demonstrate the feasibility of our proposed decomposer to be applied to any signal.
Here, we briefly tested the decomposition performance on signals that were recorded in completely different recording environments (including different measurement devices), BCI tasks, and subjects from those used for decomposer training.
We refer to this test as a cross-dataset validation.

\subsubsection{Datasets} \label{sec:datasets}
The datasets available with the Mother of all BCI Benchmarks (MOABB)~\cite{Arismunha2023} were used to build and validate the decomposer.
For detailed information on the datasets, please refer to the software website\footnote{\url{http://moabb.neurotechx.com}}.
The dataset included several datasets covering MI, P300, SSVEP, c-VEP, and resting-state paradigms.

We divided the datasets into two sets: a decomposer training set and a testing set.
The division was conducted randomly, but datasets with paradigms other than MI, SSVEP, and P300 were assigned to the training set to ensure validation of decomposition performance with standard BCI paradigms.
The training and testing sets include samples from 18 and 15 datasets, respectively.
The dataset codes are listed in the vertical axes of Figures~\ref{fig:train} and~\ref{fig:acc} for training and testing, respectively.

\subsubsection{Architecture} \label{sec:architecture}
The decomposer, consisting of eight pairs of detectors and atoms, generated eight signals.
Each detector had four one-dimensional-convolutional layers.
A layer was designed with a kernel size of 125, corresponding to a filter size of approximately 0.5~s.
The middle layers of the detector had eight channels.
The final module of the detector was a ReLU activation function, ensuring that the detector output was positive.
The atom convolution was implemented as a one-dimensional convolutional layer with a kernel size of 125, corresponding to a filter size of 0.5~s.
This layer did not have a bias term.

\subsubsection{Training} \label{sec:training}
\begin{figure}[t]
    \centering
        \includegraphics[width=.95\linewidth]{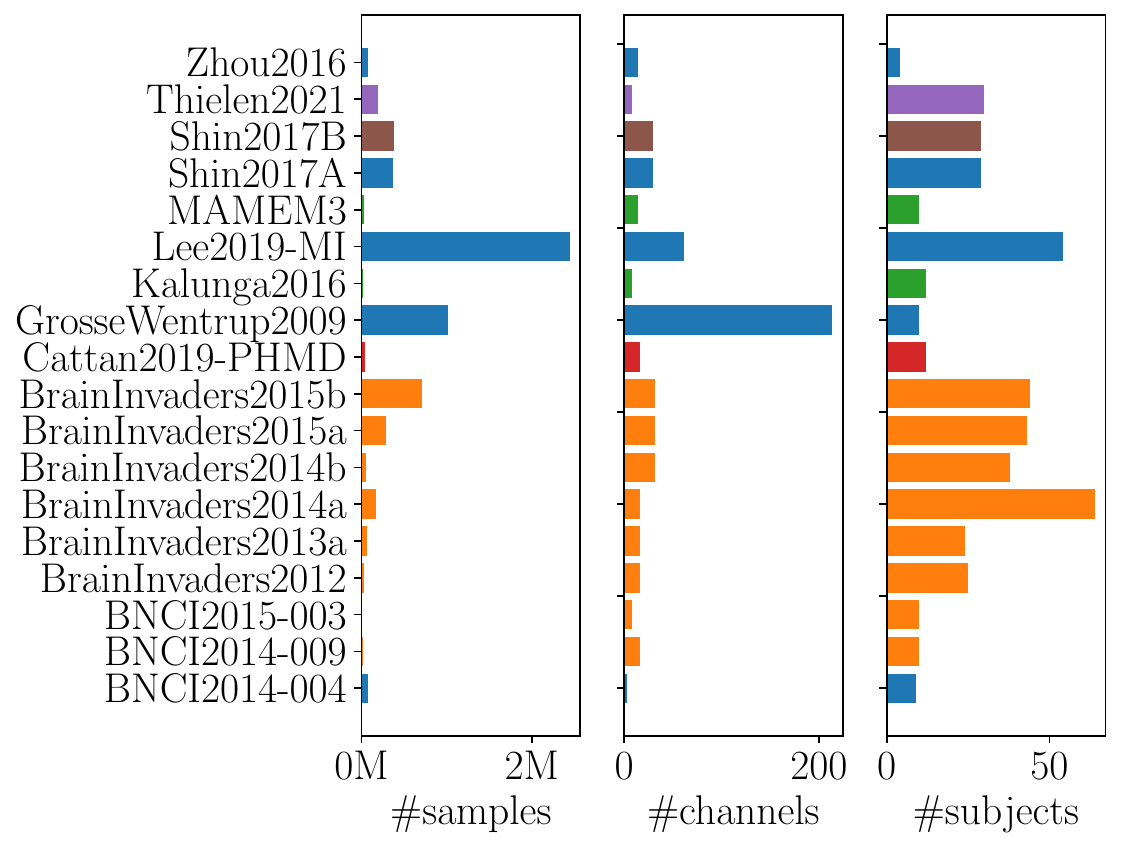}
    \caption{
        Dataset statistics for decomposer training.
        The bar colors represent the paradigms for the datasets (blue: MI, orange: P300, green: SSVEP, red: resting, purple: c-VEP, brown: arithmetic).
    }
    \label{fig:train}
\end{figure}

For organizing the decomposer training samples, we applied a common procedure for all signals in the training set.
We did not select specific electrodes; therefore, all channels were used for building the decomposer.
The signals underwent bandpass filtering using a 4th order Butterworth filter with a frequency range of 0.5 to 100~Hz.
The sampling frequency were standardized by an upsampling and downsampling scheme to 250~Hz.
We epoched the signals using a non-overlapping 3~s sliding window.
Then, the epoched signals were normalized to have a zero mean and unit variance.
This procedure did not retain the class labels, which were independently labeled for each dataset, meaning that the decomposer training was conducted in an unsupervised manner.
Finally, the total number of samples was approximately 7.48M.
The statistics (number of samples, channels, and subjects) for the training set are summarized in Figure~\ref{fig:train}.

Due to the large size of the training samples, we randomly selected 100,000 samples from the training set to accommodate our computational resources.
The optimization was performed over 100 iterations (epochs), with batches of 10,000 samples each from the selected samples.
After completing these iterations, we reselected the samples randomly and optimized the network with the new selection.
This process of random selection and optimization was repeated 140 times, resulting in a total of 14,000 epochs.
The computation took approximately five days using a single NVIDIA RTX A6000 GPU.

The network weights were initialized randomly.
The optimization process utilized an Adam optimizer with the following parameters: a learning rate of $10^{-5}$, $\beta_{1}$ of 0.5, $\beta_{2}$ of 0.999, and a weight decay of $10^{-5}$.

We monitored the power of the atoms at the beginning of every 100, applying the atom reassignment procedure described in Appendix Section~\ref{sec:atom_reassign} as needed.
The regularization coefficient $\alpha_{\text{Sparse}}$ in (\ref{eq:l_sparse}) was set to 0 for the first 4,000 epochs and adjusted to $10^{-4}$ for the remaining epochs.
Due to the significant computational resources required for training the decomposer, this coefficient was empirically tuned without additional validation.

\begin{figure}[t]
    \centering
    \includegraphics[width=.95\linewidth]{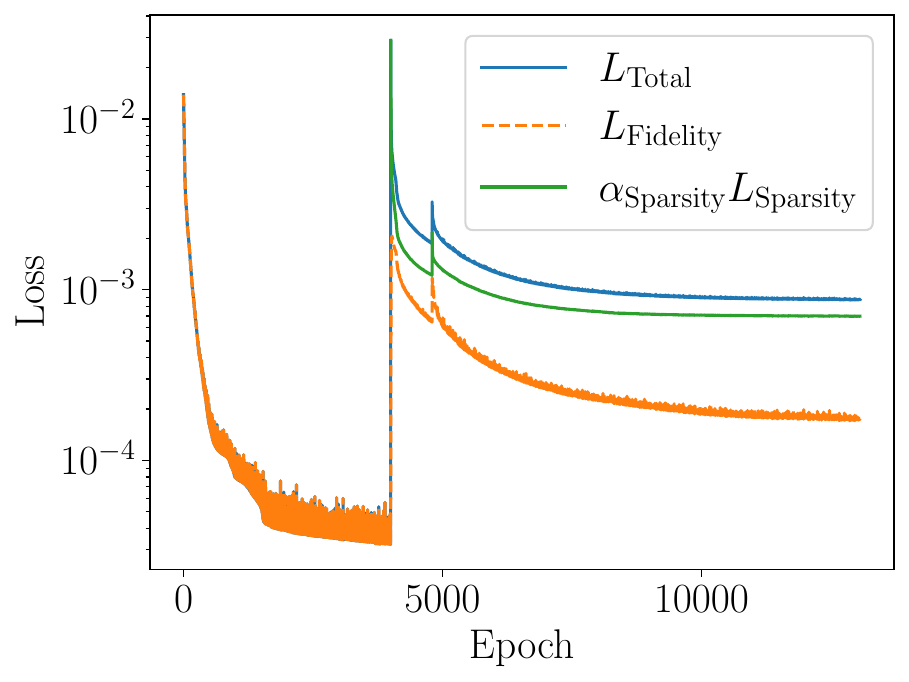}
    \caption{
        Loss for the decomposer training.
    }
    \label{fig:loss}
\end{figure}

Figure~\ref{fig:loss} illustrates the convergence of the total loss ($L_{\rm Total} = L_{\rm Fidelity} + \alpha_{\rm Sparsity} L_{\rm Sparsity}$) over the optimization epochs.
At the 4,000th epoch, the introduction of the sparsity loss (\ref{eq:l_sparse}) caused a significant fluctuated in the total loss.
Following the introduction, the loss generally decreased, with minor fluctuations due to the execution of atom replacement.
While the atom replacement theoretically does not affect the fidelity loss, it impacted the sparsity loss, which in turn affected the fidelity loss.

\begin{figure}[tp]
    \centering
    \includegraphics[width=.95\linewidth]{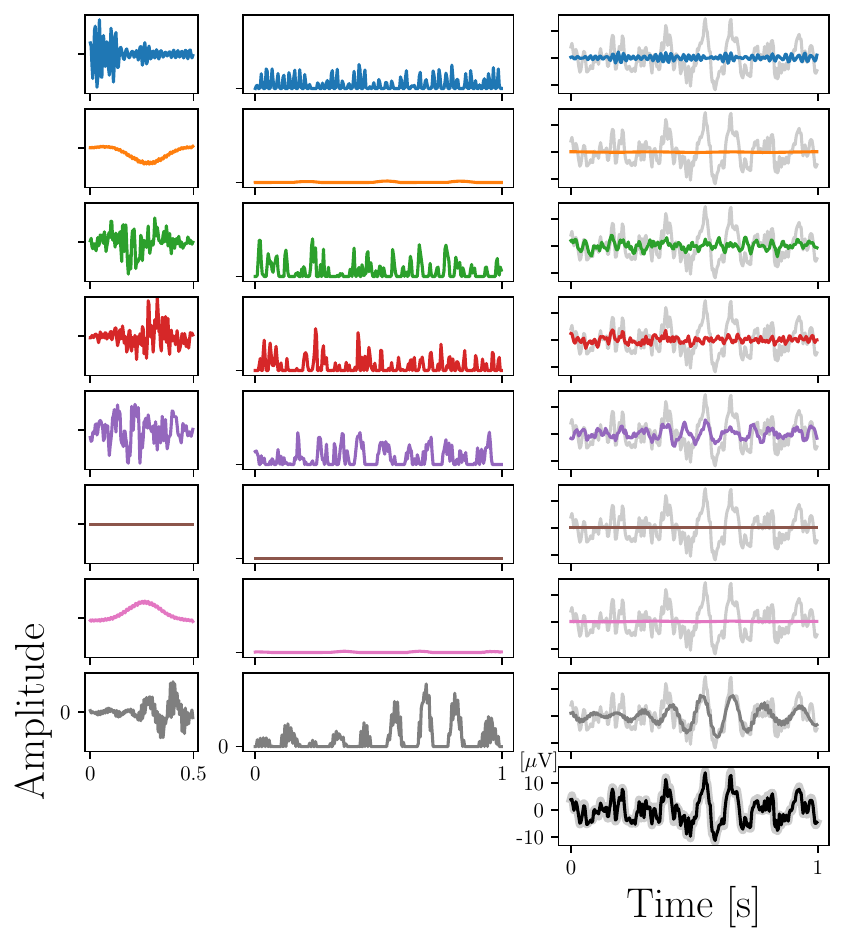}
    \caption{
        Atoms and decomposed signals.
        The top eight panels display the atoms (left), detector outputs (center), and decomposed and original signals (right).
        The bottom one shows the original (grey) and reconstructed (black) signals.
    }
    \label{fig:atoms}
\end{figure}

Figure~\ref{fig:atoms} presents the atoms in the pre-trained decomposer, along with an example of an input and output signals.
Each atom exhibits a distinct shape, seemingly representing different components within the input signal.
The reconstruction error in this example is 0.46~$\mu$V in mean absolute error (MAE), indicating successful reconstruction.

\subsubsection{Classification procedures for testing samples} \label{sec:clf}
While we epoched signals using a sliding window for the training set, the samples for the testing set were epoched according to the class labels for each dataset.
After obtaining the epoched signals, they were classified by the following procedure.

Firstly, for each dataset, we selected a single channel manually chosen for the BCI paradigm, as described in Appendix Section~\ref{sec:bci_paradigms}.
The signals were then resampled to a $250$~Hz sampling rate and bandpass filtered using a 4th order Butterworth filter with a frequency range of 0.5 to 100~Hz.
This resampling was necessary to match the sampling frequency of the decomposer and the input signal.

The resampled signals were input into the pre-trained decomposer.
The input signals were first normalized to have zero mean and unit variance.
The output decomposed signals were then inverse-normalized to match the original mean and variance.
After decomposition, the signals, which originally had a single channel, were treated as having eight channels.

These eight-channel signals were then processed through bandpass filtering, downsampling, preprocesssing systems, and classification into labels.
The specific procedures varied depending on the BCI paradigms, as described in Appendix Section~\ref{sec:bci_paradigms}.
These paradigm-specific classification procedures follow ``classic'' approaches.
In addition to the classic classifiers, we employed a deep-learning based classifier that does not require paradigm-specific preprocessing and classifier.
We used an EEGNet~\cite{Lawhern2018}-based model, specifically the EEGNet-v4 implementation from Braindecode library~\cite{Schirrmeister2017}.
Signals without any preprocessing were directly input into this model.

The classification accuracy was evaluated using 10-fold cross-validation.
It is important to note that while the decomposer had been pre-trained using different datasets, the classifiers were trained using samples from the same dataset as the testing samples.
We compared the performance of the classification with the our decomposer ({\it Proposed}) to the performance without the decomposer ({\it Original}).
In the Original condition, the procedure was the same as that for the Proposed condition, except that the decomposer was not used.

\subsubsection{Result} \label{sec:gen:result}

\begin{figure}[tp]
    \centering
    \includegraphics[width=.95\linewidth]{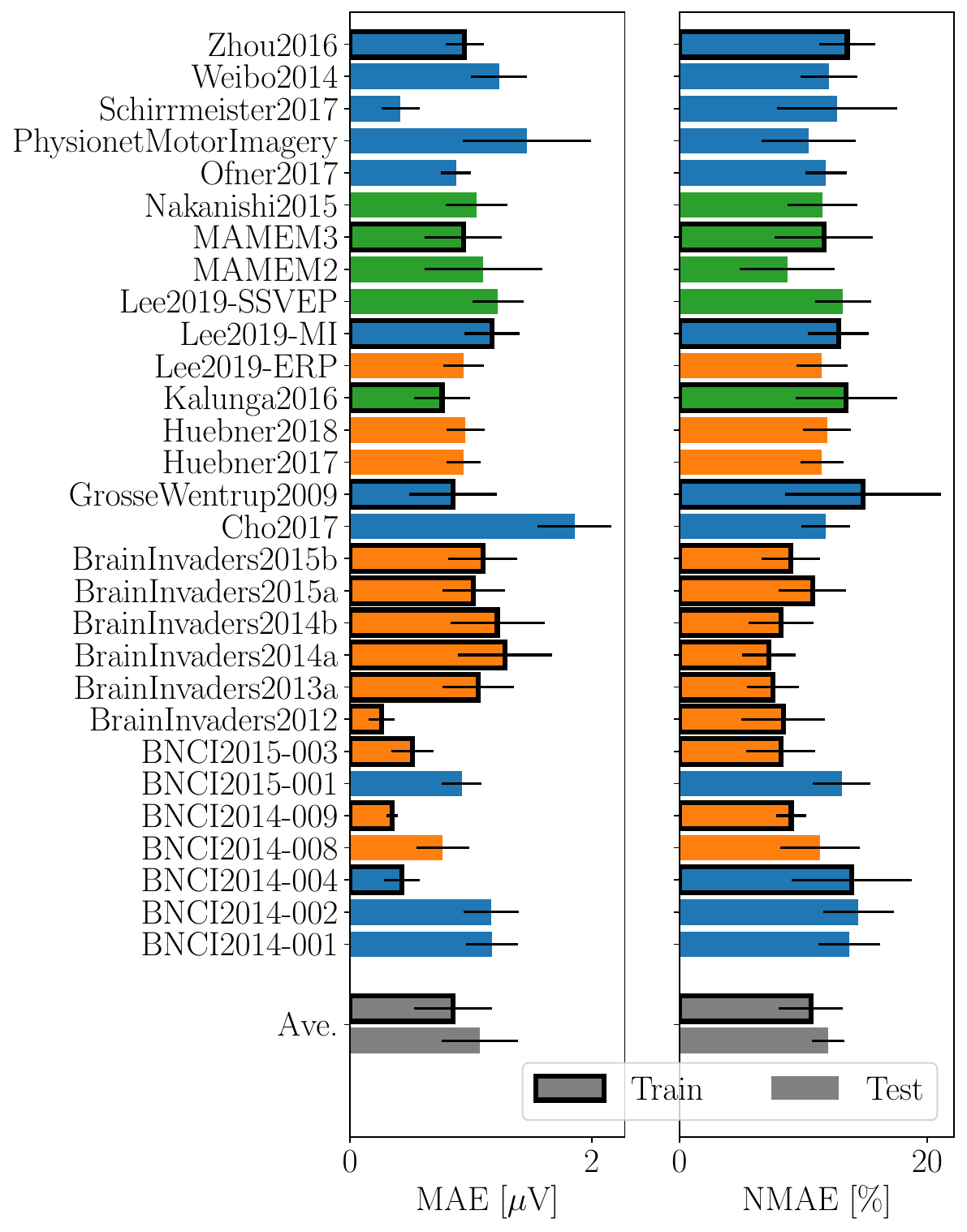}
    \caption{
        Reconstruction error for each dataset.
        The datasets with the bold edges of their bars are in the training set.
        The bar colors represent the paradigms for the datasets (blue: MI, orange: P300, green: SSVEP).
    }
    \label{fig:mae}
\end{figure}

Figure~\ref{fig:mae} shows MAEs between the original signal (input for the decomposer) and the reconstructed signal (sum of the decomposed signals).
Considering that EEG signals typically range of $\pm50$~$\mu$V, an MAE of approximately 2~$\mu$V would be relatively low.
A Mann-Whitney $U$ (MAU) test revealed no significant difference in dataset-wise MAEs (the averaged for each dataset) between the training and testing sets ($p=0.183$).
Figure~\ref{fig:mae} also shows the normalized MAE (NMAE), defined as ${\rm NMAE} = {\rm MAE} / {\rm MA}$, where ${\rm MA}$ is the mean amplitude of the original signal.
NMAE represents the ratio between the original amplitude and the reconstruction error.
The proposed decomposer achieved a ratio of approximately 10--12\%.
This metric also showed no significance difference between the training and testing sets ($p=0.214$ by an MAU test).
This lack of significant difference in the reconstruction performance suggests that overfitting is not present.

\begin{figure*}[tp]
    \centering
    \includegraphics[width=.95\linewidth]{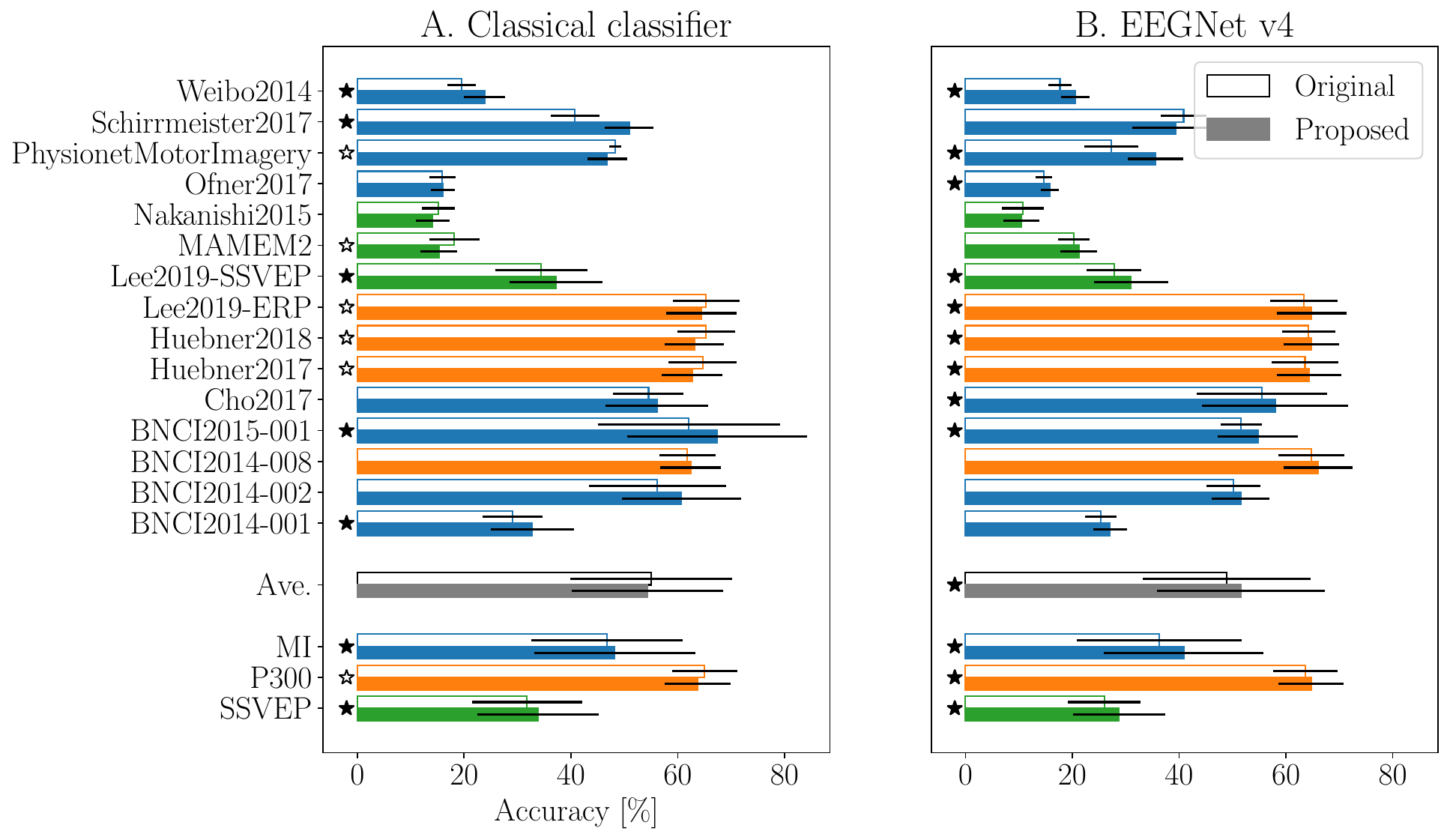}
    \caption{
        Comparison of classification accuracy rates in the cross-dataset experiment.
        We tested with classical classifiers (A) and EEGNet v4 (B) were used.
        The accuracy rates for the original (framed) and proposed (filled) conditions are shown for each dataset.
        The overall average accuracy rates across all subjects and within the same BCI paradigms are also displayed.
        Black stars indicate significant improvement by the pre-trained decomposer ($p < 0.05$), while white stars indicate significant degradation.
        The bar colors represent the paradigms for the datasets (blue: MI, orange: P300, green: SSVEP).
    }
    \label{fig:acc}
\end{figure*}

Figure~\ref{fig:acc} shows the classification accuracy for each dataset in the testing set.
The differences in accuracy rates between the proposed and original conditions were tested using an MAU test.
With the classical classifiers, as the average accuracy across all subjects showed no significant improvement (49.48\% without the decomposer vs. 50.16\% with the decomposer, $p=0.357$), we found that the decomposer did not enhance accuracy for all datasets.
However, focusing on the BCI-paradigm-wise results, the decomposer significantly improved accuracy for the MI (46.75\% vs. 48.20\%, $p=0.008$) and SSVEP (31.78\% vs. 33.83\%, $p<0.001$) paradigms, while it degraded accuracy for the P300 paradigm (65.04\% vs. 63.76\%, $p<0.001$).
This suggests that the decomposer is effectively in handling frequency components.
Decomposing time-locked components like P300 may require a different architecture, as discussed in Section~\ref{sec:mm:time-locked}.
With the EEGNet-v4 model, however, the decomposer significantly improved the classification accuracy for all paradigms (MI: 36.29\% without the decomposer vs. 40.89\% with the decomposer, $p>0.001$, P300: 63.61\% vs. 64.70\%, $p<0.001$, SSVEP: 26.02\% vs. 28.76\%, $p<0.001$), and also for the average across all subjects (42.88\% vs. 45.93\%, $p<0.001$).
Because the performance of neural network-based models depends on the sample size and the setting of training parameters, we cannot simply compare the results between the classical classifier and EEGNet-v4.
However, these improvements in the classification accuracy with the both classifiers suggest the pre-trained decomposer contributed to decoding of brain patterns.

\section{Discussion}
This paper presents a novel approach for single-channel EEG signal decomposition, based on the concept that a signal comprises multiple time-shifted and amplitude-modulated atoms~\cite{Brockmeier2016,Lewicki1999}.
Our method utilizes a neural network-based model and training data to estimate these atoms, determining their time shifts and amplitude modulations within the signal.
Consequently, we obtain decomposed signals, each associated with one of these atoms.

We evaluated the effectiveness of our decomposition method using real-world EEG data, emphasizing its contributions to neuroscience research and BCI development.
Our findings reveal that the decomposition and reconstruction process can faithfully reproduce original signals and enhance the performance of specific neuroscience and BCI tasks.
This supports our signal model based on atom convolution, indicating that the decomposed signals hold valuable information.
Additionally, our cross-dataset validation reaffirms this indication and highlights the feasibility of a pre-trained decomposer.
This pre-trained model would require no calibration and could be used in a plug-and-play manner, simplifying EEG measurement and analysis for various applications.
Further empirical research by peer researchers and developers, who can utilizing the publicly available pre-trained model, will help clarify the neuroscientific significance of our research.

One of the major issues is that our decomposer does not provide a unique decomposition.
The simultaneous optimization of the detectors and atoms is highly arbitrary, meaning that the trained detectors and atoms depend on the initial values at the start of optimization, even when the training data remains the same.
Introducing reasonable initialization for the atoms, such as using Fourier basis functions, mother wavelets, and a pre-trained model, could improve consistency for training.

Hyperparameter tuning also requires a solution.
Our current approach necessitates the tuning of network hyperparameters, particularly the number of decomposed signals and the signal length of atoms.
Expertise in the specific task domain can aid in predicting suitable parameters.
Additionally, the availability of training data is a prerequisite for our method.
Using the data, heuristic methods could be developed to facilitate parameter tuning.

Another limitation is the applicability of the pre-trained decomposer for unknown EEG patterns.
The effectiveness of the pre-trained model depends significantly on the training data.
Our decomposer produces decomposed signals through the convolution of pre-trained atoms.
Therefore, if a signal comprises atoms that are unknown or not represented in the training data, the decomposition may be incomplete.
Consequently, the pre-trained decomposer is suited for signals containing well-known and well-defined EEG components and may not be ideal for neuroscientific research aimed at discovering unknown EEG patterns.
While the pre-trained model can be valuable for preprocessing tasks like noise reduction, re-training techniques such as fine-tuning or continual learning~\cite{Parisi2019} with current datasets are recommended to enhance its applicability and accuracy.

This research introduced a novel single-channel EEG decomposition technique utilizing a convolution model with a limited number of atoms.
Our experiments provided empirical evidence supporting the applicability of this decomposition method across various EEG research scenarios.
Including the high feasibility of a pre-trained model, our approach simplifies measurement and analysis procedures, potentially advancing EEG and BCI research.

\section*{Ethics statement}
This research is not applicable.

\section*{Data and code availability}
A demo code is available in \url{https://github.com/hgshrs/detector-atom-net}.

\section*{Author Contributions}
Hiroshi Higashi is responsible for all aspects of this work, including its conception and design, analysis, and interpretation.

\section*{Funding}
This work was supported in part by the Japan Society for the Promotion of Science (JSPS) KAKENHI, grant number 22H05163 and 24K15047,
and Japan Science and Technology Agency (JST) Advanced International Collaborative Research Program (AdCORP), grant number JPMJKB2307.

\section*{Competing interests}
The authors declare no conflict of interest.


\appendix

\section{Atom reassignment} \label{sec:atom_reassign}
Let $G_{K} = \{D_{K}, \bm{a}_{K}\}$ be a pair of a detector and an atom whose generated signal converges to zero.
We can find this pair by examining the norm of the atom.
Typically, if convergence occurs, the norm approaches zero.
Suppose $G_{R} = \{D_{R}, \bm{a}_{R}\}$ is the pair selected to replace $G_{K}$.
The detector from $G_{R}$ is directly transferred to the pair $K$ as $D_{K} = D_{R}$.
For reassigning the atom, split $\bm{a}_{R}$ into two segments with the same length and allocate these to $\bm{a}_{K}$ and $\bm{a}_{R}$, respectively.
The atom $\bm{a}_{K}$ is then given the first segment, while $\bm{a}_{R}$ retains the second as
\begin{equation}
    \bm{a}_{K} =
    \begin{bmatrix}
        \bm{a}_{R}^{\prime} \\ \bm{0}
    \end{bmatrix},
    \quad
    \bm{a}_{R} =
    \begin{bmatrix}
         \bm{0} \\ \bm{a}_{R}^{\prime\prime}
    \end{bmatrix},
\end{equation}
where $\bm{a}_{R} = \left[ {\bm{a}_{R}^{\prime}}^{\top}, {\bm{a}_{R}^{\prime\prime}}^{\top} \right]^{\top}$.

This reassignment operation alters the signals generated by the generators, $G_{K}$ and $G_{R}$, but the sum, $G_{K}(\bm{x}) + G_{R}(\bm{x})$, remains equivalent to the original output of $G_{R}$.
Consequently, this operation does not affect the overall reconstructed signals or the reconstructed error.
This reassignment technique can be applied at any training epoch if zero-convergence is detected.

\section{BCI paradigms and classification procedure} \label{sec:bci_paradigms}
To evaluate the pre-trained model for signal decomposition, we assessed classification performance using standard BCI paradigms.
The classification procedures were implemented using the straightforward, standard methods provided in MOABB, as the primary aim of this evaluation was to validate whether the decomposer effectively decomposes an input signal into meaningful components.
Detailed descriptions of these procedures can be found in the software's manual~\cite{Arismunha2023}.
Below, we provide a brief overview of the procedures for each BCI paradigm, with all settings maintained at the software's default configurations.

\subsection{Motor imagery}
For classification, a signal recorded from a specific electrode was used.
Given that each dataset may have varying electrode configurations, the electrode was selected from a predefined list of candidates: C3, C4, CP3, CP4, C1, C2, CP1, CP2, C5, C6, CP5, CP6, Cz, CPz, Pz, and Fz.
The order of these candidates reflects their priority for selection; if a dataset lacks the C3 channel, C4 is chosen instead.

Signal samples were created by epoching the signal from the task onset to a dataset-specific end time, without applying baseline correction.
The signals were filtered within a frequency range of 8 to 32~Hz using a 4th order Butterworth filter.
The common spatial pattern (CSP) method~\cite{blankertz2008} was employed on the sample signals, reducing the number of channels to a minimum of eight.
The logarithm of the powers for the channels of was input into a classifier based on Fisher's linear discriminant analysis (LDA)~\cite{Friedman1989}.

\subsection{P300}
A single channel was selected from the options of Cz, CPz, FCz, Pz, C3, C4, CP1, CP2, P3, and P4 for classification.
Signal samples were created by epoching the signal from the task onset to a dataset-specific end time, without applying baseline correction.
The signals were filtered within a frequency range of 1 to 24~Hz using a 4th order Butterworth filter.
The vectorized form of the multi-channel signals was then input into a classifier based on LDA~\cite{Krusienski2006,Blankertz2011}.

\subsection{SSVEP}
A single channel was selected from the options of Pz, POz, Pz, O1, O2, PO3, and PO4 for classification.
Signal samples were created by epoching the signal from the task onset to a dataset-specific end time.
The signals were bandpass filtered between 7 to 45~Hz using a 4th order Butterworth filter.
For classification, the canonical correlation analysis (CCA)-based method~\cite{Lin2006} was employed.
This involved computing the canonical correlation between each sample and a set of sinusoidal and cosinusoidal waves for each flickering frequency and its three harmonics.
A sample was classified into the flickering frequency label corresponding to the set that yielded the highest canonical correlation.
This classification procedure was entirely unsupervised.

\end{document}